\documentclass[aps,nofootinbib,twocolumn,groupedaddress,floatfix,eqsecnum,preprintnumbers]{revtex4}
\usepackage[dvips]{color}
\usepackage{color}
\usepackage{graphicx}
\usepackage{epsfig}
\usepackage{epstopdf}
\usepackage{rotating}
\usepackage[mathscr]{eucal}
\usepackage{amsmath,amssymb,bm,graphics}
\usepackage[pdftex,letterpaper=true]{hyperref}
\usepackage{yfonts}
\usepackage{bm}

\def\coeff#1#2{{\textstyle \frac{#1}{#2}}}
\let\mathbf=\bm

\begin{document}

\def\Nfour{\mathcal N\,{=}\,4}
\def\Ntwo{\mathcal N\,{=}\,2}
\def\Nc{N_{\rm c}}
\def\Nf{N_{\rm f}}
\def\x{\mathbf x}
\def\q{\mathbf q}
\def\f{\mathbf f}
\def\v{\mathbf v}
\def\C{\mathcal C}
\def\vs{v_{\rm s}}
\def\S{\mathcal S}
\def\half{{\textstyle \frac 12}}
\def\twothirds{{\textstyle \frac 23}}
\def\third{{\textstyle \frac 13}}
\def\t{\mathbf{t}}
\def\T{\mathcal {T}}
\def\O{\mathcal{O}}
\def\E{\mathcal{E}}
\def\p{\mathcal{P}}
\def\H{\mathcal{H}}
\def\uh{u_h}
\def\R{\ell}
\def\Ro{\chi}
\def\del{\nabla}
\def\eps{\hat \epsilon}
\def\nn{\nonumber}
\def\K{\mathcal K}
\def\inf{\epsilon}
\def\cs{c_{\rm s}}
\def\A{\mathcal{A}}
\def\e{{e}}
\def\x{{\mathbf x}}
\def\r{{\mathbf r}}
\def\freq{\omega_0}
\def\rr{{\xi}}
\def\uo{{u_*}}
\def\u{{\mathcal U}}
\def\G{\mathcal{G}}
\def\Deltax{\Delta x_{\rm max}}

\title
{Synchrotron radiation in strongly coupled conformal field theories}

\author{Christiana~Athanasiou$^1$}
\author{Paul~M.~Chesler$^1$}
\author{Hong~Liu$^1$}
\author{Dominik~Nickel$^2$}
\author{Krishna~Rajagopal$^1$}

\affiliation
   {$^1$Department of Physics, Massachusetts Institute of Technology, Cambridge, MA 02139, USA\\
%
$^2$Institute for Nuclear Theory, University of Washington, Seattle, WA 98195}

\date{January 21, 2010}

\begin{abstract} 
Using gauge/gravity duality, we compute the energy density and angular distribution of the power radiated by a quark 
undergoing circular motion in strongly coupled ${\cal N}=4$ supersymmetric Yang-Mills (SYM) theory. We compare the strong coupling results to those at weak coupling, and find the same angular distribution of radiated power, up to an overall prefactor.  In both regimes, the angular distribution is in fact similar to that of synchrotron radiation produced by an electron in circular motion in classical electrodynamics:
the quark emits radiation in a narrow beam along its velocity vector
with a characteristic opening angle $\alpha \sim 1/\gamma$.  
To an observer far away from the quark, the emitted radiation appears as a short periodic burst, just like the light from a lighthouse does to a ship at sea.  Our strong coupling results are valid for any strongly coupled conformal field theory with a dual classical gravity description.
\end{abstract} 

\preprint{INT-PUB-10-004}
\preprint{MIT-CTP-4112}


\maketitle
\iftrue
\def\thefootnote{\fnsymbol{footnote}}
\def\thefootnote{\arabic{footnote}}
\fi

\section{Introduction and Outlook}

The emission and propagation of radiation is an interesting and natural
topic of study in any field theory.
For localized sources, one is typically 
interested in the angular distribution and spectrum of radiation
at spatial infinity and in deciphering how information about the source 
is encoded in it.  This topic has been studied for more than a century within the 
realm of classical electrodynamics (for a textbook treatment, see \cite{Jackson}).  
It is much less studied in realms where quantum effects can be important, as for example
in a nonabelian gauge theory in the limit of strong coupling.  
This has largely been
due to the absence of strong coupling methods for studying real-time dynamics.  
With the discovery of gauge/gravity duality, however,
the dynamics of quantum fields in 
certain strongly coupled nonabelian gauge theories can be mapped onto
the dynamics of classical fields living in a higher dimensional spacetime \cite{Maldacena:1997re}.
Because of this,  strongly coupled real-time phenomena in theories 
with gravitational duals have become accessible to reliable calculation.

To date, much of the analysis of real-time phenomena in gauge theories with dual gravitational descriptions has focused on systems at nonzero temperature.  This is because these strongly coupled plasmas are more similar to the strongly coupled quark-gluon plasma of QCD being produced and studied in heavy ion collision experiments at the Relativistic Heavy Ion 
Collider~\cite{Adcox:2004mh,Back:2004je,Arsene:2004fa,Adams:2005dq} than the vacua of these gauge theories are to the vacuum of QCD.   Asking how a test quark moving along some trajectory with nonzero acceleration radiates in vacuum in a strongly coupled nonabelian gauge theory is a simpler question than many of those that have already been  studied at nonzero temperature.   We shall answer this question for the specific case of a test quark in circular motion in the vacuum of ${\cal N}=4$ supersymmetric Yang-Mills (SYM) theory.  

${\cal N}=4$ SYM theory is a conformal nonabelian gauge theory, specified by two parameters: the number of colors $N_c$ and the  't Hooft coupling $\lambda\equiv g^2 N_c$ with $g^2$ the gauge coupling.  We shall work in the large $N_c$ limit throughout.  Since the theory is conformal, we can specify $\lambda$.  If we choose $\lambda\rightarrow 0$, we can study the radiation using methods similar to those used in classical electrodynamics.
If we choose $\lambda\rightarrow \infty$, the dual gravitational description of this strongly coupled nonabelian quantum field theory becomes classical.  Specifically, it becomes the classical gravity limit of Type IIB string theory in the AdS$_5 \times$S$_5$ spacetime.  The geometry of the five-sphere will play no role in our computation --- indeed it can be replaced by any compact five-dimensional space    $X_5$.  This means that our strong coupling results will be valid for all conformal quantum field theories with a dual classical gravity description --- since conformality of the quantum theory maps onto the presence of an AdS$_5$ spacetime in the gravitational description.\footnote{Our results are valid for all strongly coupled conformal quantum field theories with a dual classical gravity description as long as we express our results in terms of the 't Hooft coupling $\lambda$ in each of the conformal field theories.  The relations between this parameter, $N_c$,  and the parameters that specify the gravitational physics in the AdS$_5$ space --- namely the string coupling and the dimensionless ratio of the square of the AdS curvature and the string tension --- will be different in different conformal field theories, since these relations do depend on the geometry of $X_5$.}

Because the vacuum of ${\cal N}=4$ SYM theory is not similar to that of our world, studying the radiation of a test quark in circular motion in this theory does not have direct phenomenological motivation.   However, we shall find that the angular distribution of the radiation is very similar to that of classic synchrotron radiation: when the quark is moving along its circle with an increasingly  relativistic velocity, the radiation is produced in an increasingly tightly collimated beam --- a beam which, if Fourier analyzed, contains radiation at increasingly short wavelengths and high frequencies.   This is interesting first in its own terms and second with a view to extending the calculation to nonzero temperature.  First, from the point of view of the quantum field theory
we do not understand why the qualitative properties of the radiation are so similar at $\lambda\rightarrow\infty$ and
$\lambda\rightarrow 0$.  Why should the pattern of radiation when what is radiated is nonabelian and strongly coupled be similar to that seen in classical electrodynamics?  In this sense, our results come as something of a surprise to us. Second, our results in this formal setting nevertheless open the way to a new means of modelling jet quenching in heavy ion collisions.
If we were to add a nonzero temperature to our calculation, we could watch the tightly collimated beam of synchrotron radiation interact with the strongly coupled plasma that would then be present.  The beam of radiation should be slowed down from the speed of light to the speed of sound and should ultimately thermalize, and it would be possible to study how the length- and time-scales for these processes depend on the wavelength and frequency of the beam.  


We shall consider a test quark moving on a circle of radius $R_0$ 
with constant angular velocity $\omega_0$.   In spherical coordinates $\{r,\theta,\varphi\}$ we can take the quark's trajectory to be given by
\begin{equation}
r=R_0,\quad \theta=\frac{\pi}{2},\quad \varphi=\omega_0 t\ .
\label{CircularMotion}
\end{equation}
By a test quark we mean a test charge in the fundamental representation of the $SU(N_c)$ gauge group.
The total power radiated by this test quark moving in the vacuum of ${\cal N}=4$ SYM theory in the large-$N_c$ and strong coupling limit has been computed previously by Mikhailov in Ref.~\cite{Mikhailov:2003er}, who finds
\begin{equation}
P=\frac{\sqrt{\lambda}}{2\pi} a^2\ ,
\label{MikhailovPower}
\end{equation}
where $a$ is the quark's proper acceleration, given by 
\begin{equation}
a=\gamma^2 v \omega_0 \ ,
\end{equation}
where the quark has speed
\begin{equation}
v=R_0\omega_0
\end{equation}
and where 
\begin{equation}
\gamma\equiv \frac{1}{\sqrt{1-v^2}}
\end{equation}
is the standard Lorentz factor.
Mikhailov observed that upon making the substitution
\begin{equation}
\sqrt{\lambda}\leftrightarrow \frac{e^2}{3} 
\label{MikhailovSubstitution}
\end{equation}
the result (\ref{MikhailovPower}) becomes identical to the Larmor result in classical electrodynamics.
Furthermore, Mikhailov showed that the power radiated by a test quark moving along an 
{\it arbitrary} trajectory is identical to what Li\'enard calculated in 1898~\cite{Lienard} --- see \cite{Jackson} for a textbook treatment --- upon making the substitution (\ref{MikhailovSubstitution}).    We shall go beyond Mikhailov's result by computing the distribution in space and time of the power radiated by the quark in circular motion.  We shall not attempt Mikhailov's generalization to an arbitrary trajectory.

Before leaping to the conclusion that Mikhailov's result suggests  that the ${\cal N}=4$ SYM synchrotron radiation pattern will be similar to that in classical electrodynamics, consider the results in the literature that push one's intuition in the opposite direction.
One interesting study related to zero temperature radiation 
was performed in Ref.~\cite{Hofman:2008ar}, where 
decays of off-shell bosons in strongly coupled holographic
conformal field theories were analyzed. (One can think of the off-shell boson as the virtual photon produced in electron-positron annihilation, with that photon coupling to matter in a strongly coupled conformal field theory instead of to quarks in QCD.)  
It was shown in Ref.~\cite{Hofman:2008ar} 
that there is no correlation between the boson's 
spin and the angular distribution of power radiated through the sphere at infinity.
In other words, if one prepares a state containing an off-shell boson with definite spin,
in the rest frame of the boson the event averaged angular distribution of power at infinity will always be isotropic.  
This stands in stark contrast to similar weak coupling calculations in QCD, where the boson's spin 
imprints a distinct ``antenna" pattern in the distribution of power at infinity 
\cite{Basham:1978zq, Basham:1978bw, Basham:1977iq}.  
Similar behavior regarding the isotropization of radiation at strong 
coupling was also reported in Refs.~\cite{Hatta:2008tx,Chesler:2008wd}.
A natural question to ask when considering the results of Ref.~\cite{Hofman:2008ar}
is where does the isotropization of radiation come from?  
Is isotropization a characteristic of the particular initial states studied in Ref.~\cite{Hofman:2008ar} or is it a natural process which happens during the propagation of strongly coupled radiation?
Indeed, it has been suggested 
in Ref.~\cite{Hatta:2008tx} that the mechanism responsible for the isotropization is that 
parton branching is not suppressed at strong coupling and, correspondingly,
successive branchings can scramble any initially preferred 
direction in the radiation as it propagates out to spatial infinity.
Regardless of mechanism, these calculations provide an example in which the pattern of radiation is very different in the strongly coupled ${\cal N}=4$ SYM theory to that in weakly coupled QCD or QED.    If the intuition about the emission and propagation of radiation derived from these calculations applies to radiation from an accelerated test charge, then, we should expect that in our strongly coupled nonabelian theory this radiation should isotropize and ``branch'', populating longer and longer wavelength modes as it propagates.   We find nothing like this.  The straightforward guess based upon Mikhailov's result --- namely that synchrotron radiation in strongly coupled ${\cal N}=4$ SYM theory should look like that in classical electrodynamics --- turns out to be close to correct.


We shall analyze the angular distribution of power radiated by a rotating quark in ${\cal N}=4$ SYM theory at both weak and strong coupling. Our weak coupling results are valid only in this theory.  At strong coupling,
we obtain an analytic expression for the energy density which is valid at
all distances from the quark and for all holographic conformal field theories.  
At weak coupling our analysis reduces to that of 
synchrotron radiation in classical electrodynamics
with an additional contribution coming from a scalar field \cite{Chesler:2009yg}.
At strong coupling, gauge/gravity duality maps the problem onto that of
the linear gravitational response due to the presence of a string.

\begin{figure}[t]
\includegraphics[scale=0.44]{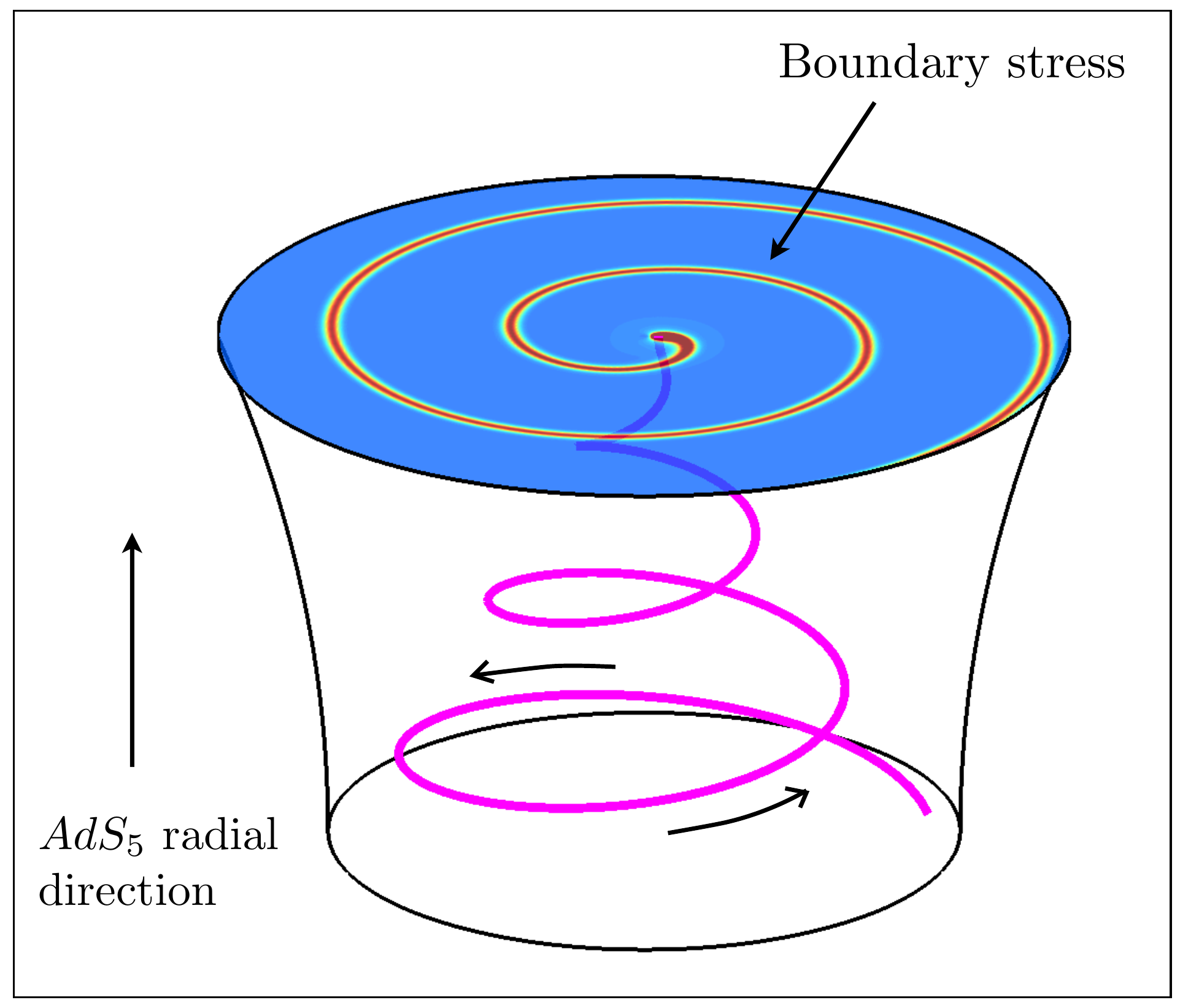}
\caption
    {\label{bulk2boundary}
   A cartoon of the gravitational description of synchrotron radiation at strong 
   coupling.  The arena where the gravitational dynamics takes place is the $5d$ geometry
   of AdS$_5$.  A string resides in the geometry and an endpoint
   of the string is attached to the $4d$ boundary of $AdS_5$, which is where the dual 
   quantum field theory lives.   The trajectory of the endpoint of the string 
   corresponds to the trajectory of the dual quark.  Demanding that the endpoint rotates 
   results in the string rotating and coiling around on itself as it extends in the $AdS_5$ radial direction.
   The presence of the string in turn perturbs the geometry and the near-boundary
   perturbation in the geometry induces a $4d$ stress tensor on the boundary.  The induced stress 
   has the interpretation as the expectation value of the stress tensor in the dual quantum field theory.
    }
\end{figure}

Fig.~\ref{bulk2boundary} shows a cartoon of the classical $5d$ dual gravitational description.
The five dimensional geometry of AdS$_5$,
which is the arena where the dual gravitational dynamics takes place, has a four-dimensional
boundary whose geometry is that of ordinary Minkowski space.   Ending at the boundary 
is a classical string whose endpoint follows a trajectory that corresponds to the trajectory of the quark in the dual field theory.  
The condition that the endpoint rotates about a specific (Minkowski space) axis results in the entire string rotating 
about the same axis with the string coiling around on itself over and over.  
In turn, the rotating motion of the string creates gravitational radiation 
which propagates up to the boundary of the geometry.
Just like electromagnetic fields induce surface currents on conductors,
the gravitational disturbance near the boundary induces a $4d$ stress tensor on the boundary \cite{Brown:1992br, deHaro:2000xn}.
As shown in Fig.~\ref{bulk2boundary}, the induced $4d$ stress tensor inherits the coiled structure of the rotating string.
Via the standard gauge/gravity dictionary \cite{Maldacena:1997re, Witten:1998qj}, 
the induced $4d$ stress tensor is the expectation value of the stress tensor in the dual quantum field theory.   So, by doing a classical gravitational calculation in five dimensions (which happens to be somewhat analogous to a classical electromagnetic calculation) we can compute the pattern of radiation in the boundary quantum field theory, including all quantum effects and working in a strong coupling regime in which quantum effects can be expected to dominate.

At both weak and strong coupling we find that the radiation pattern at 
infinity qualitatively resembles that of synchrotron radiation in classical electrodynamics.
And, at both weak and strong coupling, for the case of relativistic circular motion we find the characteristic pattern of a  lighthouse-like beam propagating out to infinity without broadening.
As stated above, at weak coupling the difference between our results and those of classical electrodynamics simply amount to the presence of an additional 
contribution coming from a scalar field.  Remarkably, up to a normalization, we find that both the total power
and the time-averaged angular distribution of power are the same at weak coupling and strong coupling.

As the radiation is far from isotropic at 
infinity, this problem provides a concrete example demonstrating
that radiation need not isotropize in a nonabelian gauge theory
at strong coupling.  And, the width of the outgoing pulse of radiation does not broaden as the pulse propagates outward, meaning that as it propagates out to infinity the radiation remains at the (short) wavelengths and (high) frequencies at which it was emitted.

We close this introduction with a look ahead at extending this calculation to nonzero temperature $T$.   The authors of Ref. \cite{Fadafan:2008bq} have shown that the power that it takes to move the quark in a circle is given by the vacuum result (\ref{MikhailovPower}) even for a test quark moving through the strongly coupled plasma present at $T\neq 0$ as long as 
\begin{equation}
\omega_0^2 \gamma^3 \gg \pi^2 T^2\ .
\label{StirringCriterion}
\end{equation}
In the opposite regime, where $\omega_0^2 \gamma^3 \ll \pi^2 T^2$, the power it takes to move the quark in a circle through the plasma is the same as it would be to move the quark in a straight line with speed $v$.  This power  has been computed using gauge/gravity duality in Refs.~\cite{Herzog:2006gh,Gubser:2006bz}, and corresponds to pushing the quark against a drag force.  The analogue of our calculation --- namely following where the dissipated or ``radiated'' power goes --- was done, again using gauge/gravity duality, in
Refs.~\cite{Gubser:2007xz,Chesler:2007an, Gubser:2007ga,Chesler:2007sv}. 
There, it was demonstrated that the test quark plowing through the strongly coupled plasma in a straight line excites hydrodynamic modes of the plasma, trailing a diffusion wake and --- if the motion is supersonic --- creating a Mach cone.  
It is a generic feature of a nonzero temperature plasma, at weak coupling or at strong coupling, that any power dumped into it in some localized fashion must eventually take the form of 
hydrodynamic excitations \cite{Chesler:2007sv}.  These considerations and our results motivate the extension of the present calculation to nonzero temperature, in particular in the regime (\ref{StirringCriterion}).  At these high velocities or low temperatures, we know that (\ref{MikhailovPower}) is satisfied and, given our experience at zero temperature in this paper, we therefore expect that at length scales that are small compared to $1/T$ the rotating quark radiates a narrow beam of synchrotron radiation, even at strong coupling.   If (\ref{StirringCriterion}) is satisfied,  the wavelengths characterizing the pulse of radiation heading outward into the plasma are much shorter than $1/T$.  But, we also know on general grounds that this narrow pulse of beamed radiation must first reach local thermal equilibrium, likely converting into an outgoing hydrodynamic wave moving at the speed of sound, and must then ultimately dissipate and thermalize completely.
Watching these processes occur may give insight into jet quenching in the strongly coupled quark-gluon plasma of QCD.

The outline of the rest of our paper is simple: we do the weak coupling calculation in Section II and use gauge/gravity duality to do the strong coupling calculation in Section III.  In Section IV we discuss the results and conclude.

\section{Weak coupling calculation}
\label{weakcoupling}

In the limit of weak coupling, the radiation produced by an
accelerated test charge can be analyzed using classical field theory.
For example, for the case of an electron undergoing synchrotron 
motion in QED, the appropriate classical effective description is simply that
of classical electrodynamics.  For the case of a test quark (specifically, an infinitely massive particle from the ${\cal N}=2$ hypermultiplet that is in the fundamental representation of the $SU(N_c)$ gauge group) moving through
the vacuum of ${\cal N}=4$ SYM theory, 
more fields can be excited and the appropriate effective classical description is more complicated.  
The Lagrangian for this theory can be found, for example, in Ref.~\cite{Chesler:2006gr}.
As can readily be verified by inspecting the field theory Lagrangian,
in the limit of asymptotically weak coupling and large test quark mass
the fields that can be excited by the test quark are the non-Abelian 
gauge field and an adjoint representation scalar field.  Interactions with all other fields are 
either suppressed by an inverse power of the quark mass or a positive power of
 the `t Hooft coupling $\lambda$.  
Moreover, in the limit of asymptotically weak coupling the gauge 
and scalar fields satisfy decoupled linear equations of motion. 
 
Following Ref.~\cite{Chesler:2009yg}, instead of dealing with $\Nc^2$ decoupled 
components of the gauge field and $\Nc^2$ decoupled components of 
the adjoint scalar field, we consider the case of a single 
$U(1)$ gauge field and a single real scalar field with appropriate effective
couplings tailored to compensate for the fact that the actual theory contains adjoint
representation fields.
The effective couplings can be determined by 
 computing the vector and scalar contributions to the Coulomb 
field of a quark in the effective description and matching onto the full theory.
In the large $\Nc$ limit the classical effective Lagrangian reads \cite{Chesler:2009yg}
\begin{equation}
\label{eq:classicalL}
\mathcal L_{\rm classical} = -\frac{1}{4} F_{\mu \nu}F^{\mu \nu} - J \cdot A  - \frac{1}{2} (\partial \chi)^2 - \rho \chi \, ,
\end{equation}
where $F^{\mu \nu}$ is the field strength corresponding to the $U(1)$ gauge field $A^\mu$, 
$\chi$ is the scalar field, and the sources corresponding to the test quark are given by
\begin{eqnarray}
\label{eq:sources}
\rho &=& e_{\rm eff} \sqrt{1- v^2} \delta^3(\bm r - \bm r_{\rm quark})\, , \\
 J^{\mu} &=& e_{\rm eff} \frac{d r^\mu_{\rm quark}}{dt} \delta^3(\bm r - \bm r_{\rm quark})\, ,
\end{eqnarray}
with $r_{\rm quark}^{\mu}$ the trajectory of the test quark, 
 $v = \dot {\bm r}_{\rm quark}^2$ the velocity of the test quark,
and where it turns out that both effective couplings take on the same value
\begin{equation}
e_{\rm eff}^2 = g^2 \frac{N_c^2-1}{2 N_c} \simeq \frac{1}{2} \lambda\, .
\label{e_eff}
\end{equation}
In classical electrodynamics, $\rho=0$ and,  in the source for the vector field $J^\mu$, 
$e_{\rm eff}$ is just $e$.  We shall take the trajectory of the test quark to be (\ref{CircularMotion}) throughout.

\subsection{Solutions to the equations of motion and the angular distribution of power}

Comprehensive analyses of the radiation coming from the gauge field can be found in classical electrodynamics textbooks (see for example \cite{Jackson}).  However, because the case of scalar radiation is less well known and also for the sake of completeness, we present a brief analysis of the radiation for both fields.  We shall linger in our discussion on those aspects which will also be of value in the analysis of our strong coupling results.

Choosing the Lorentz gauge $\partial \cdot A =0$, the equations of motion 
for the fields read
\begin{equation}
- \partial^2 A^\mu = J^\mu\,, \ \ \ -\partial^2 \chi = \rho\,.
\end{equation}
These equations are solved by
\begin{subequations}
\label{grnsfcnsols}
\begin{eqnarray}
A^\mu(t,\bm r) &=& \int d^4r' G(t{-}t',\bm r{-} \bm r') J^\mu(t',\bm r')\,, \\
\chi(t,\bm r) &=& \int d^4r' G(t{-}t',\bm r{-} \bm r') \rho(t',\bm r')\,,
\end{eqnarray}
\end{subequations}
where 
\begin{equation}
G(t{-}t',\bm r{-} \bm r') = \frac{\theta(t{-}t')}{4 \pi |\bm r {-} \bm r'|} \delta(t -t'-|\bm r {-} \bm r'|)
\end{equation} 
is the retarded Greens function of $-\partial^2$, and $(t',\bm r')$ is the point of emission.
After using delta functions to carry out the $r'^\mu$ integrations, we arrive at the expressions
\begin{subequations}
\label{classicalsols}
\begin{eqnarray}
\label{Asol}
A^\mu &=& \frac{e_{\rm eff}}{4 \pi r } \frac{1}{\Xi(t',\bm r)}\frac{d r^\mu_{\rm quark}}{dt'} \Big |_{t' = t_{\rm ret}}\,, 
 \ \ \ \ \ \ \ 
\\ \label{chisol}
\chi &=& \frac{e_{\rm eff}}{4 \pi  r}  \frac{1}{\Xi(t',\bm r)} \frac{1}{\gamma} \Big |_{t' = t_{\rm ret}}\,, 
 \ \ \ \ \ \ \ 
\end{eqnarray}
\end{subequations}
where
\begin{equation}
\label{xieq}
\Xi(t',\bm r) \equiv  \frac{| \bm r {-} \bm r_{\rm quark}(t') |    -    \bm r \cdot \dot {\bm r}_{\rm quark}(t')}{r}\,,
\end{equation}
$\gamma = 1/\sqrt{1- v^2}$, $r=|\bm r|$, and the retarded time $t_{\rm ret}$ is the solution to
\begin{equation}
\label{tret0}
t - t_{\rm ret} - | \bm r - \bm r_{\rm quark}(t_{\rm ret})| = 0\,.
\end{equation}
$(t-t_{\rm ret})$ is the light travel time between $\bm r_{\rm quark}(t_{\rm ret})$ and $\bm r$.
Because the motion of the quark is periodic, $\bm r_{\rm quark}(t_{\rm ret})$ 
is a periodic function of its argument with period $2\pi/\freq$.  Eq.~(\ref{tret0}) then implies that if 
$t$ is shifted by $2\pi/\freq$, $t_{\rm ret}$ shifts by the same amount so that $(t-t_{\rm ret})$ is left unchanged.  Because the motion of the quark is circular, Eq.~(\ref{tret0}) also implies that a change in the position of the quark corresponding to a shift in its azimuthal angle by $\delta\varphi$ is equivalent to a shift in both $t$ and $t_{\rm ret}$ by $\delta\varphi/\freq$. 

\begin{figure}
\includegraphics[scale=0.43]{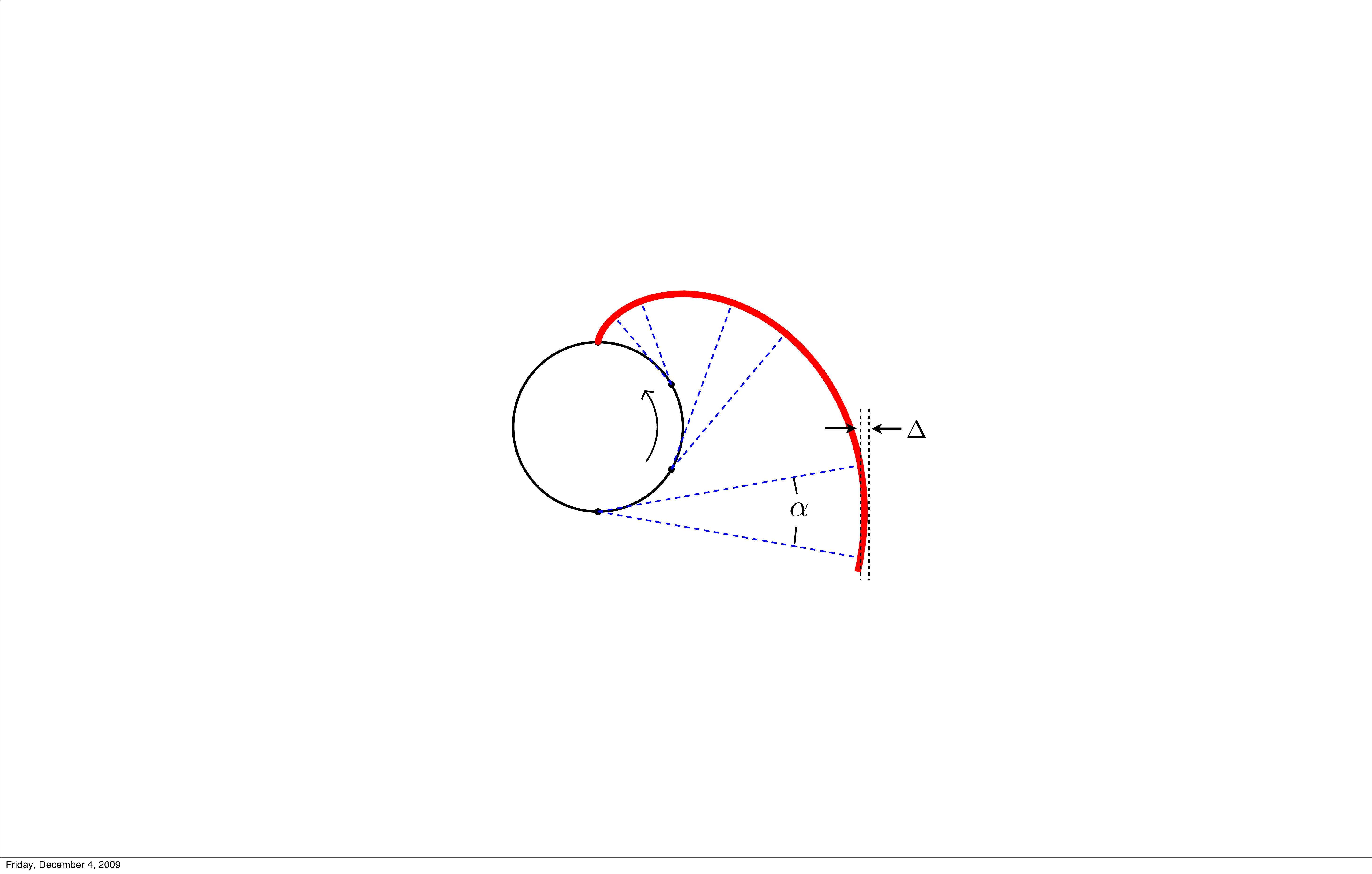}
\caption
    {\label{grnsfcnsol}
   A sketch of the solutions
   (\ref{classicalsols}).  As the quark moves along its circular trajectory,
   it emits radiation in a narrow cone of angular width $\alpha \sim 1/\gamma$ 
   in the direction of its velocity vector.  The diagram is a snapshot at the time when the quark
   is at the top of the circle.  The red spiral shows where the radiation emitted at earlier times is located at the time of the snapshot. The width
   $\Delta$ of the spiral scales like $\Delta \sim 1/\gamma^3$, as explained in Fig.~\protect{\ref{widthdiagram}}.
   }
\end{figure}

\begin{figure}[t]
\includegraphics[scale=0.50]{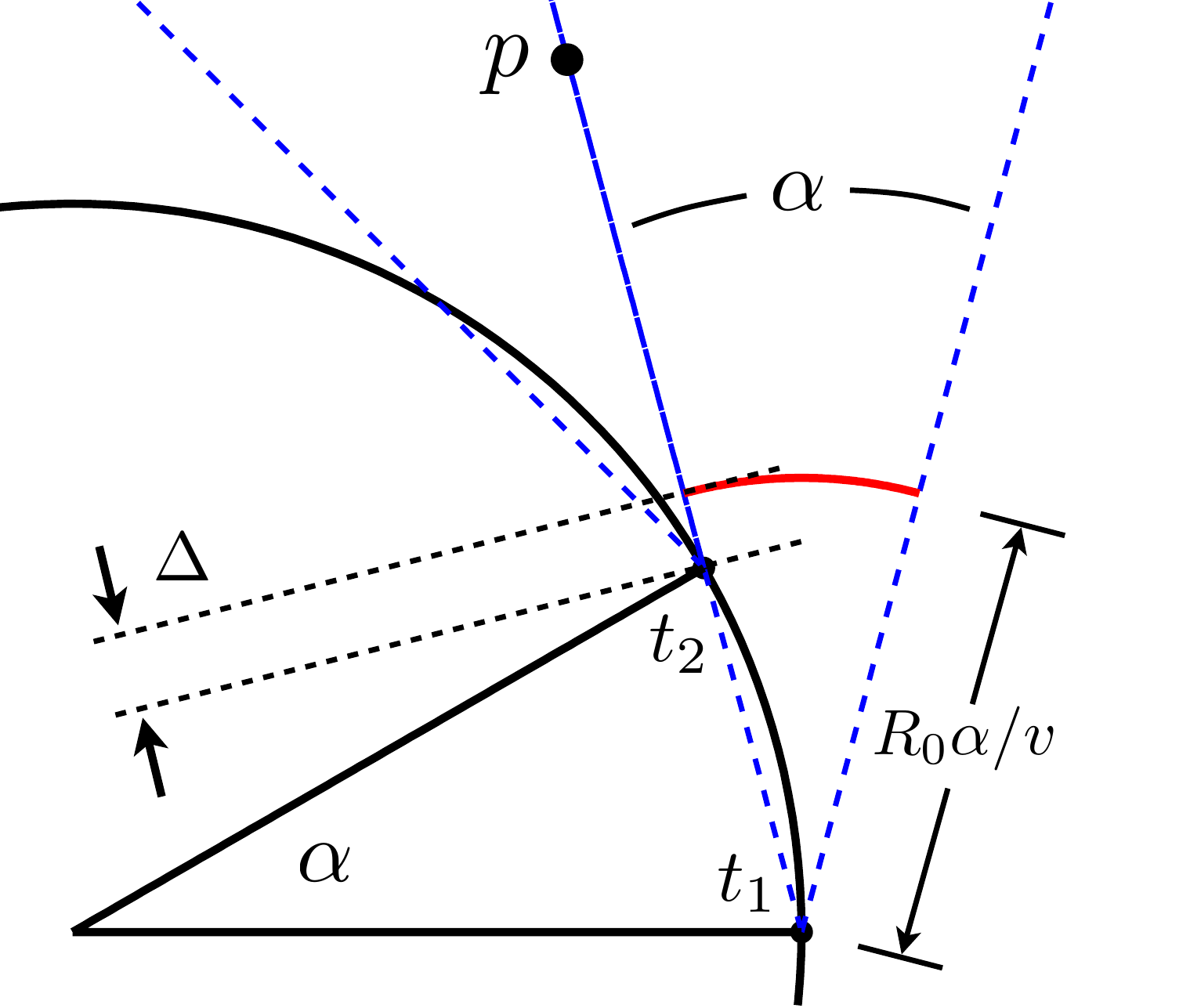}
\caption
    {\label{widthdiagram}
   A close-up illustration of the emission of radiation 
   at two times $t_1$ and $t_2$.  The radiation is emitted in the direction
   of the quark's velocity vector, within a cone of angular width $\alpha$.  An observer at the point
   $p$ is illuminated by a pulse of radiation of duration $\Delta t \sim R_0 \alpha$
   and of spatial thickness $\Delta$.
   The leading edge of the pulse observed at $p$ is emitted at $t_1$ and the trailing edge
   observed at $p$ is emitted at $t_2$.  At time $t_2$ the radiation emitted at 
   $t_1$, denoted by the solid red line, has traveled a distance $R_0 \alpha/v$ towards
   $p$.   The chordal distance between the two emission points is $R_0 \alpha$ in the 
   $\alpha \to 0$ limit.  The width $\Delta$ is therefore $\Delta = R_0 \alpha(1/v - 1).$
    }
\end{figure}

We now describe the qualitative behavior of the solutions
(\ref{classicalsols}).  Fig.~\ref{grnsfcnsol} shows a pictorial
representation of the solutions (\ref{classicalsols}).
As the quark moves along its circular trajectory, it emits radiation
in the direction of its velocity in a narrow cone of angular width $\alpha$.
From the solutions (\ref{classicalsols}) it can be shown that in the large $\gamma$ limit
$\alpha$ scales like
$\alpha \sim 1/\gamma$ for both the scalar and vector radiation.
The cone of radiation emitted at each time propagates outwards at the speed 
of light.   At any one moment in time, the radiation emitted at all times in the past forms a spiral, as illustrated by the red spiral sketched in Fig.~\ref{grnsfcnsol}.   Clearly, the radial width $\Delta$ of the spiral must go to zero as $\alpha\to 0$, namely as $\gamma\to \infty$.  As we illustrate in Fig.~\ref{widthdiagram}, in the large $\gamma$ limit the radial width scales like $\Delta \sim 1/\gamma^3$.
%
%
%
To understand the scaling of $\Delta$, consider an observer at the point $p$
in Fig.~\ref{widthdiagram}.  As the quark moves along its trajectory, 
an observer at $p$ will see a short pulse of radiation
of duration $\Delta t \sim R_0\alpha/v$.  The leading edge of the pulse
will be emitted by the quark at time $t_1$ and the trailing edge will be emitted 
at time $t_2 = t_1 + R_0 \alpha/v$.  At time $t_2$ the radiation emitted at 
$t_1$ will have traveled a distance $R_0 \alpha/v$ towards $p$.  Moreover, the chordal 
distance between the two emission points is approximately $R_0 \alpha$
when $\alpha$ is small.  It therefore follows that the spatial thickness $\Delta$ of the pulse 
observed at $p$ scales like 
\begin{equation}
\label{Deltascaling}
\Delta \sim (R_0 \alpha/v - R_0 \alpha) \sim R_0 \alpha/\gamma^2 \sim R_0/\gamma^3\,.
\end{equation}
We note that the fact that $\Delta$ and $\alpha$ scale with different powers of $\gamma$ is a consequence of the fact that 
the radiation is emitted in the direction of the quark's velocity.  If the radiation were emitted
in any other direction, $\Delta$ and $\alpha$ would have the same $\gamma$ scaling. 
Because we will re-use these arguments in the analysis of our strong coupling results, it is important to note that they are purely geometrical, relying only on the fact that the radiation travels at the speed of light and the fact that the outward going pulse of radiation propagates without broadening, maintaining the spiral shape whose origin we have illustrated.    

In addition to determining how the pulse widths $\alpha$ and $\Delta$ are related, these geometrical considerations in fact specify the location of the spiral of radiation precisely.  For a quark moving in a circle of radius $R_0$ in the $(x,y)$ plane, at the time at which the quark is located at $(x,y)=(R_0,0)$ the spiral of energy density radiated at all times in the past is centered on a curve in the $z=0$ plane specified by
\begin{equation}
\label{curve}
x = R_0 \! \left [ \cos \rho {+} { \frac{\rho}{v} } \sin \rho \right],
 \ y = R_0 \! \left [-\sin \rho {+}  { \frac{\rho}{v} }\cos \rho \right]\,, 
 \end{equation}
where $\rho$ is a parameter running from 0 to $\infty$.   At large $r$, this spiral can equivalently be written 
\begin{equation}
\varphi(r,t) = \frac{\pi}{2} +\freq t - \frac{r v}{R_0}+{\cal O}(1/r)
\label{curveLargeR}
\end{equation}
in spherical coordinates, where we have also added the $t$-dependence.

Recall that the $\gamma$-scaling of $\Delta$ depended on the fact that the radiation is emitted tangentially, in the direction of the velocity vector of the quark. The same is true for the shape of the spiral.  For example, if we repeat the geometrical analysis for a hypothetical case in which the quark emits a narrow cone of radiation in the $\hat r $ direction, perpendicular to its direction of motion, the energy density would be centered on a different spiral specified by
\begin{equation}
x =  R_0 \left [ 1  + { \frac{\rho}{v} } \right]  \cos \rho\,,
 \ y = - R_0 \left [ 1  + { \frac{\rho}{v} } \right]  \sin \rho\,.
\label{WrongSpiral}
\end{equation}
This spiral, including its $t$-dependence, is given in spherical coordinates by
\begin{equation}
\varphi(r,t)=v +\freq t - \frac{r v}{R_0} \,,
\end{equation}
which differs from the correct spiral of Eqs.~(\ref{curve}) and (\ref{curveLargeR}) by a phase 
shift at large $r$. 
If the quark emits radiation perpendicular to its direction of motion,
a distant 
point $p$ would be illuminated by light emitted when the quark was at the closest point on its circle to $p$, whereas for the correct spiral (\ref{curve}) the light seen at $p$ was emitted one quarter of a period earlier but had travelled $R_0$ farther, leading to the phase shift of $(-\pi/2+v)$.  In the near zone, (\ref{WrongSpiral}) differs from the correct spiral (\ref{curve}) in shape, not just by a phase shift.

In Section IV, we shall compare our results at strong coupling to those we have obtained in this Section via geometrical arguments built upon the fact that the outward going pulse of radiation propagates without spreading.

We now analyze the far zone behavior of the fields and compute the time-averaged
angular distribution of power.  In particular, we wish to compute the angular distribution
of power radiated through the sphere at $r = \infty$ averaged over one cycle of the quark's motion.  
Denoting the energy flux by $\bm S$, 
the time averaged angular distribution of power is given by
\begin{equation}
\label{power0}
\frac{d P}{d \Omega} = \frac{\freq}{2 \pi} \lim_{r \to \infty} r^2 \int_{0}^{\frac{2 \pi}{\freq}} dt_{\rm ret} \frac{\partial t}{\partial t_{\rm ret}} \, \hat r \cdot \bm S\,,
\end{equation} 
where, from Eqs.~(\ref{tret0}) and (\ref{xieq}), 
\begin{equation}
\frac{\partial t}{\partial t_{\rm ret}} = \frac{r\Xi}{t-t_{\rm ret}}\,.
\label{xiRelation}
\end{equation}
(Technically, the integration in Eq.~(\ref{power0}) should run from
  $t_{\rm ret} = c$ to $t_{\rm ret} = c + 2 \pi /\freq$ where
  $c\sim -r$. This is simply because in the far-zone limit
  the retarded time always scales like $t_{\rm ret} \sim - r$.  However,
  as the motion of the quark is periodic one can always shift the integration
  variable so that the integration runs from $0$ to $2 \pi /\freq$.)

For the simple case of scalar or electromagnetic fields, the computation of the energy flux
directly from the fields is relatively simple.  However, in the gravitational setting discussed below, the
computation of the energy flux requires solving three differential equations whereas
the computation of the energy density requires only solving one.  Therefore, if possible, it is 
useful to extract the angular distribution of power from the energy density alone. 

For a conformal theory at zero temperature the far-zone energy density and flux associated 
with a localized source must take the form
\begin{equation}
\label{farzonelimits}
\mathcal E(t,\bm r) = \frac{\varepsilon(t_{\rm ret},\theta,\varphi)}{r^2}\,, \ \ \bm S(t,\bm r) = \frac{s(t_{\rm ret},\theta,\varphi)}{r^2} \hat r\,,
\end{equation}
up to $O(1/r^3)$ corrections.  Moreover, using the continuity equation 
and the fact that $\partial t_{\rm ret}/\partial t = -\partial t_{\rm ret}/\partial r$
in the far zone, it is easy to see that we must have $\varepsilon = s$.  
In terms of $\varepsilon$, the time-averaged
angular distribution of power reads
\begin{equation}
\label{power1}
\frac{d P}{d \Omega} = \frac{\freq}{2 \pi} \int_{0}^{\frac{2 \pi}{\freq}} dt_{\rm ret} \frac{\partial t}{\partial t_{\rm ret}} \varepsilon\,.
\end{equation} 
We see from Eq.~(\ref{power1}) that we can extract the angular distribution of the radiated power at infinity by analyzing the asymptotic behavior of the energy density.

The energy density associated with the vector field
is $\mathcal E_{\rm vector} = \frac{1}{2} (\bm E^2 + \bm B^2)$ where $\bm E$ and $\bm B$ are the electric and 
magnetic fields. The energy density associated with the scalar is 
$\mathcal E_{\rm scalar} = \frac{1}{2} \left [ (\partial_t \chi)^2 +  (\del \chi)^2 \right ]$.  Evaluating
these quantities for the solutions (\ref{classicalsols}) in the $r\rightarrow\infty$ limit yields
\begin{subequations}
\label{WeakCouplingEps}
\begin{align}
\mathcal{E}_{\rm vector}=&
\frac{e_{\rm{eff}}^2 v^2 \omega _0^2 }{64 \pi ^2 r^2 \xi^6}
\Big[
3
+2 v^2
+\left(1-2 v^2\right) \cos 2 \theta
\nonumber\\&
-
8 v \sin\theta \sin(\varphi-\omega_0t_{\rm ret})
\nonumber\\&
-
2 \sin^2\theta \cos \left(2(\varphi-\omega_0t_{\rm ret})\right)
\Big]
\,,\\
\mathcal{E}_{\rm scalar}=&
\frac{e_{\rm{eff}}^2 v^2 \omega _0^2 }{16 \pi ^2 \gamma^2 r^2 \xi^6}
\,
\sin^2\theta \cos^2 (\varphi-\omega_0t_{\rm ret})\,,
\, 
\end{align}
\end{subequations}
up to subleading corrections proportional to $1/r^3$, where we have defined
\begin{equation}
\xi \equiv 1 -v\sin\theta\sin(\varphi-\omega_0t_{\rm ret})\, ,
\label{xiInftyDefn}
\end{equation}
which is the $r\to \infty$ limit of $\Xi$:
\begin{equation}
\Xi=\xi + {\cal O}(1/r)\, .
\label{XiExpansion}
\end{equation}
Note that the energy densities (\ref{WeakCouplingEps}) are proportional to $1/\xi^6$ and therefore  have maxima in the $\theta=\pi/2$ plane where $\xi$ has minima; at large $r$, these occur on the spiral (\ref{curveLargeR}).

Noting from (\ref{tret0}) that $(t-t_{\rm ret})\sim r$ in the $r\to\infty$ limit, we see  
from (\ref{xiRelation}) and (\ref{XiExpansion})  that  $\partial t/\partial t_{\rm ret}=\xi$ in the $r\to \infty$ limit.
We can now evaluate the expression (\ref{power1}) for the  time-averaged angular distribution of power, yielding
\begin{subequations}
\label{weakpower}
\begin{align}
\label{vector}
\frac{d P_{\rm vector}}{d\Omega}&=\frac{v^2 \freq^2 e_{\rm eff}^2}{128 \pi^2} \frac{8 - 4 \sin^2 \theta - v^2 \left(1 + 3 v^2 \right ) \sin^4 \theta}{\left ( 1 - v^2 \sin^2 \theta \right )^{7/2}} ,
   \\
\label{scalar}
\frac{d P_{\rm scalar}}{d\Omega}
&= \frac{v^2 \freq^2 e_{\rm eff}^2}{128 \pi^2 \gamma^2} \frac{\sin^2 \theta \left ( 4 + v^2 \sin^2 \theta \right )}{\left ( 1 - v^2 \sin^2 \theta \right )^{7/2}} 
\,.
\end{align}
\end{subequations}
Eqs.~(\ref{weakpower})  show that the power radiated in vectors and in scalars are both 
focused about $\theta = \pi/2$ with a characteristic angular 
width $\delta \theta \propto 1/\gamma$.  (Since $1-v^2 \sin^2 \theta = 1/\gamma^2$ at $\theta=\pi/2$ and  $1-v^2 \sin^2\theta \simeq (1+c)/\gamma^2$ at $\theta=\pi/2 - c/\gamma$ for any constant $c$ when $\gamma\gg 1$.)
Adding the contributions (\ref{vector}) and (\ref{scalar}), we find that the time-averaged angular distribution of power radiated in vectors plus scalars is
\begin{equation}\label{VectorPlusScalar}
\frac{d P_{\rm total}}{d\Omega}
= \frac{v^2 \freq^2 e_{\rm eff}^2}{32 \pi^2} \frac{ \left ( 2 + v^2 \sin^2 \theta \right )}{\left ( 1 - v^2 \sin^2 \theta \right )^{5/2}} \ .
\end{equation}

Integrating Eqs.~(\ref{weakpower}) over all solid angles we find the power radiated in
vectors and scalars
\begin{subequations}
\label{classicalpower}
\begin{eqnarray}
\label{vectorpower}
P_{\rm vector} &=& \frac{2}{3} \frac{e_{\rm eff}^2}{4 \pi} a^2\,,
\\ \label{scalarpower}
P_{\rm scalar} &=& \frac{1}{3} \frac{e_{\rm eff}^2}{4 \pi} a^2\,,
\end{eqnarray}
\end{subequations}
where $a = v \freq/(1-v^2)$ is the proper acceleration of the test quark.
With $e_{\rm eff}=e$, Eq.~(\ref{vectorpower}) becomes the well known Larmor formula 
for the power radiated by an accelerated charge in classical electrodynamics.  Evidently,
at weak coupling two times as much energy is radiated into the
vector mode as the scalar mode.   We now see that Mikhailov's strong coupling result (\ref{MikhailovPower}) is equivalent to the result for weakly coupled ${\cal N}=4$ SYM theory upon making the substitution
\begin{equation}
\sqrt{\lambda} \leftrightarrow \frac{1}{2}  e_{\rm eff}^2 \ ,
\label{CorrectSubstitution}
\end{equation} 
with the difference between this result and (\ref{MikhailovSubstitution}) --- which was obtained by comparing the strong coupling ${\cal N}=4$ SYM result to classical electrodynamics --- coming from the fact that in weakly coupled ${\cal N}=4$ SYM theory the rotating test quark radiates scalars as well as vectors.

\section{Gravitational Description and Strong Coupling Calculation}


Gauge/gravity duality~\cite{Maldacena:1997re} provides a description of the (strongly coupled; quantum mechanical) vacuum of
${\cal N}=4$ SYM theory in $N_c\rightarrow\infty$ and $\lambda\rightarrow\infty$ limit in terms of classical gravity in the AdS$_5\times S_5$ spacetime.  As we discussed in Section I, generalizing from ${\cal N}=4$ SYM to any other conformal quantum field theory that has a classical gravity dual involves replacing the $S_5$ by some other compact five-dimensional space.  In what follows, we shall never need to specify this five-dimensional space, meaning that all our strong coupling results will be valid for any large-$N_c$ conformal field theories with gravitational duals.

The metric of the AdS$_5$ spacetime is
\begin{equation}
\label{metric}
    ds^2 = \frac{L^2}{u^2}
    \left [-f \, dt^2 + d \bm r^2 + \frac{du^2}{f} \right ] \,,
\end{equation}
where $L$ is the $AdS_5$ curvature radius, $u$ is the (inverse) $AdS_5$ radial coordinate
with the boundary at $u = 0$, and $f = 1$.
(We have kept the factor of $f$ in the metric in order to generalize some of our calculations to nonzero temperature, where 
\begin{equation}
\label{fBH}
f=1-\frac{u^4}{u_h^4}
\end{equation}
with $u_h=1/(\pi T)$ and $T$  the temperature of the field theory state.)
A single test
quark in the fundamental representation in the field theory corresponds in the gravitational description to an
open string with one endpoint attached to the boundary at $u = 0$.
Taking the $N_c\rightarrow \infty$ and $\lambda\rightarrow\infty$ limits eliminates fluctuations of the string and makes the probability for loops to break off the string vanish.  Upon taking these limits, the string is therefore a classical object.
The presence of the classical string hanging ``down'' into the five-dimensional bulk geometry perturbs that geometry via Einstein's equations, and the behavior of the metric perturbation near the boundary encodes the change in the boundary stress tensor. 
This is illustrated in Fig.~\ref{bulk2boundary}, and in this Section we shall turn this cartoon into calculation.

\subsection{The rotating string}
\label{RotatingStringSection}

We begin by describing the rotating string.
The metric perturbations that this string creates are small in the large $N_c$ limit, small enough that we can neglect their back reaction on the shape of the string itself.  We can therefore begin by finding the shape of the rotating string in AdS$_5$, in the absence of any metric perturbations.
The dynamics of classical strings are governed by the Nambu-Goto action
\begin{equation}
S_{\rm NG} = - T_0 \int d \tau \,d \sigma \sqrt{-g}\,,
\label{SNG}
\end{equation}
where $T_0 = \sqrt{\lambda}/{2 \pi L^2}$ is the string tension, $\sigma
$ and $\tau$ are
the coordinates on the worldsheet of the string, and $g= {\rm det}\,
g_{ab}$ where $g_{ab}$ is
the induced worldsheet metric.  The string profile is determined by a
set of embedding
functions $X^M(\tau,\sigma)$ that specify where in the spacetime
described by the metric $G_{MN}$ the point $(\tau,\sigma)$ on the
string worldsheet is located. The induced world sheet metric is given
in terms of these functions by
\begin{equation}
g_{ab} = \partial_a X \cdot \partial_b X\,,
\end{equation}
where $a$ and $b$ each run over $(\tau,\sigma)$. For the determinant we obtain
\begin{equation}
-g= (\partial_\tau X \cdot \partial_\sigma X)^2 - (\partial_\tau X)^2
(\partial_\sigma X)^2\,.
\end{equation}

We choose worldsheet coordinates $\tau = t$ and $\sigma = u$.
As we are interested in quarks which rotate at constant frequency
$\freq$ about the $\hat z$ axis, we parametrize the string embedding
functions
via
\begin{equation}
X^{M}(t,u) = (t,\bm r_s(t,u),u)\,,
\label{Xstring}
\end{equation}
where in spherical coordinates $\{r,\theta,\varphi\}$ the three-vector
$\bm r_s$ is given by
\begin{equation}
\bm r_s(t,u) \equiv \left (R(u), {\textstyle \frac{\pi}{2}},\phi(u) +
\freq t \right ).
\label{StringWorldsheetParametrization}
\end{equation}
The function $\phi(u)$ describes how the purple string in Fig.~
\ref{bulk2boundary} winds in azimuthal angle as a function of
``depth'' in the AdS$_5$ radial direction, whose coordinate is $u$.
The function $R(u)$ describes how the string in Fig.~
\ref{bulk2boundary} spreads outward as a function of depth.  Because
we are describing a situation in which the test quark at the boundary
has been rotating on its circle for all time, the purple string in
Fig.~\ref{bulk2boundary} is rotating at all depths with the same
angular frequency $\omega_0$, meaning that the two functions $R(u)$
and $\phi(u)$ in the parametrization
(\ref{StringWorldsheetParametrization}) fully specify the shape of the
rotating string at all times.
With this parameterization, the Nambu-Goto action reads
\begin{equation}
\label{NB2}
S_{\rm NG} = - \frac{\sqrt{\lambda}}{{2 \pi}} \int dt \,du \, \mathcal
L\,,
\end{equation}
where
\begin{equation}
\mathcal L = \frac{\sqrt{(1-\freq^2 R^2 )(1 + R'^2) + R^2 \phi'^2}}
{u^2}\,.
\label{NGLagrangian}
\end{equation}
To determine the shape of the string in Fig.~\ref{bulk2boundary}, we
must extremize (\ref{NB2}).

The equations of motion for $\phi(u)$ and $R(u)$ follow from
extremizing the
Nambu-Goto action (\ref{NB2}).  One constant of motion can be obtained
by noting
that the action is independent of $\phi(u)$.  This implies that
\begin{equation}
\Pi \equiv  - \frac{\partial \mathcal L}{\partial \phi'}
\label{PiDef}
\end{equation}
is constant.  Here and throughout, by $'$ we mean $\partial/\partial u
$. The minus sign on the right-hand side of (\ref{PiDef})  is there in order to make $\Pi$ positive for positive $\omega_0$.
 In terms of $\Pi$, the equation of motion for $\phi'(u)$
reads
\begin{equation}
\label{phieq}
\phi'^2 = \frac{u^4 \Pi^2 (1 - \freq^2 R^2)(1 + R'^2)}{R^2(R^2 - u^4
\Pi^2)}\,.
\end{equation}
The equation of motion for $R(u)$ is given by
\begin{equation}
\frac{\partial \mathcal L}{\partial R} - \frac{\partial}{\partial u}
\frac{\partial \mathcal L}{\partial R'}  = 0\,.
\end{equation}
Taking the above partial derivatives of $\mathcal L$ and then
eliminating $\phi$ derivatives via Eq.~(\ref{phieq}),
we obtain the following equation of motion:
\begin{align}
\label{rhoeq}
R'' + \frac{R (u + 2 R R') (1 + R'^2)}{u (u^4 \Pi^2 - R^2)}
+ \frac{1 + R'^2}{R(1 - \freq^2 R^2)} = 0.
\end{align}

As a second order differential equation, one might expect that
Eq.~(\ref{rhoeq}) requires two initial conditions for the
specification of a unique solution.  However, it was shown in Ref.~\cite{Fadafan:2008bq} that
this is not the case --- together, Eqs.~(\ref{phieq}) and (\ref{rhoeq}) imply that
all required initial data for (\ref{rhoeq}) is fixed in terms of $\Pi$ and $\freq$.  To see how this comes about note that
Eq.~(\ref{rhoeq}) is singular when $1 - \freq^2 R^2 = 0$ or when $R^2 - u^4 \Pi^2 = 0$.
In order to maintain the positivity of the right hand side of
Eq.~(\ref{phieq}), $1 - \freq^2 R^2$ and $R^2 - u^4 \Pi^2$
must vanish at the same point. This happens where 
\begin{align}
u =  u_c \equiv \frac{1}{\sqrt{\Pi\freq}}, \ \ R  = R_c \equiv \frac{1}{\freq}.
\end{align} 
We can then solve (\ref{rhoeq}) in the vicinity of $u=u_c$ by expanding $R(u)$ as
\begin{equation}
\label{ucseries}
R(u) = R_c + R'_c (u - u_c) + \frac{1}{2} R''_c (u - u_c)^2 + \dots,
\end{equation}
where $R^{(n)}_c \equiv \partial_u^n R|_{u {=} u_c}$.
The second and third terms in (\ref{rhoeq}) are divergent at $u=u_c$ while the $R''$ term is finite there.  
This means that when we substitute (\ref{ucseries}) into (\ref{rhoeq}) 
and collect powers of $(u-u_c)$, we obtain an equation for $R'_c$ that does not involve 
$R''_c$, meaning that the equation of motion itself determines $R'_c$.  
A short calculation yields
\begin{equation}
\label{Rseries}
R'_c = \frac{1}{2}\left [ \sqrt{\frac{\freq}{\Pi} + 4} -  \sqrt{\frac{\freq}{\Pi}} \right ].
\end{equation}
Therefore, both $R_c$ and $R_c'$ are  
determined by $\freq$ and $\Pi$ and consequently, 
for given $\freq$ and $\Pi$ there is only one solution to Eq.~(\ref{rhoeq}).
From the perspective of the boundary field theory
this had to happen: for a given angular frequency $\freq$, the
radiation produced by the rotating quark 
can only depend on the radius $R_0$ of the quark's trajectory 
(which as we will see below can be related to $\Pi$).

With the above point in mind, the unique solution to Eq.~(\ref{rhoeq})
is given by%
\footnote
  {
  One way to derive the solution (\ref{RSolution}) is simply to solve
  Eq.~(\ref{rhoeq}) with the series expansion (\ref{Rseries}) and recognize the 
  resulting series as that of (\ref{RSolution}).
  }
\begin{equation}
R(u) = \sqrt{v^2 \gamma^2 u^2 + R_0^2}\,,
\label{RSolution}
\end{equation}
where $v = R_0 \freq$ is the velocity of the quark
and $\gamma = 1/\sqrt{1-v^2}$.  The velocity of the quark is related to
$\Pi$ via
\begin{equation}
\Pi = v^2 \gamma^4 \freq\,.
\label{PIandV}
\end{equation}
With $R(u)$ given by (\ref{RSolution}), it is easy to integrate
Eq.~(\ref{phieq}).  Taking
the negative root of Eq.~(\ref{phieq}) so that the string trails its $u=0$ 
endpoint, we find
\begin{equation}
\phi(u) = - u \gamma \freq + \arctan\left(u \gamma \freq\right)
\label{phiSolution}
\end{equation}
and thus 
 \begin{equation}
\varphi(t,u) =\omega_0(t - \gamma u  ) + \arctan\left(u \gamma \freq\right) \,.
\label{varphiS}
\end{equation}
We thus have an analytic description of the shape of the rotating string.

We can obtain the profile of the rotating string projected onto the plane at the boundary by using (\ref{RSolution}) to eliminate $u$ from (\ref{varphiS}), yielding 
 \begin{equation}
 \varphi (t,r) = \omega_0 t - \sqrt{{r^2 \over R_0^2} -1} +  \arctan\left(\sqrt{{r^2 \over R_0^2} -1} \right) \,.
 \label{ProjectedString}
 \end{equation}
In the limit of $r \gg R_0$, we have 
 \begin{equation} \label{spiS}
 \varphi (t,r) = \omega_0 t + {\pi \over 2} - {r \over R_0}  + {\cal O}(1/r) \,.
 \end{equation}
Equation~\eqref{spiS} describes a spiral whose neighboring circles (at fixed $t$) are separated by $2\pi R_0$.  Note that the curve (\ref{spiS}) is the same spiral as (\ref{curveLargeR}) when the velocity of the quark $v\to 1$.   Indeed, it can also be shown that
at all radii  (\ref{ProjectedString}) is the same spiral as (\ref{curve}) with $v=1$. We conclude that for $v\to 1$ the rotating string hanging down into the AdS$_5$  geometry is lined up precisely below the spiral in the boundary energy density in the limit of weak coupling and, as we shall see below, in the limit of strong coupling.\footnote
{
This identification suggests interesting possibilities at nonzero temperature.
Whereas at $T=0$ the rotating string has a profile $R(u)$ such that it extends out to infinite $r$, when $T\neq 0$ the rotating string only extends out to some finite maximum radius $r_{\rm max}$ that depends on $\omega_0$ and $v$~\cite{Fadafan:2008bq}.  
The fact that at zero temperature we can identify the the spiral of the rotating string with the spiral of energy density in the boundary quantum field theory suggests that, at least at $v\to 1$, the maximum radius out to which the rotating string reaches may be some  length scale related to how far the beam of synchrotron radiation can travel through the hot strongly coupled plasma before it thermalizes.  Interestingly, if one takes the $v\to 1$
at fixed $R_0$, one finds $r_{\rm max}\to\infty$~\cite{Fadafan:2008bq}. 
}
At velocities less than $1$, the shape of the rotating string is not related to the shape of the spiral of energy density, at least not directly by projection.  This is perhaps not surprising, as it is only in the $v\rightarrow 1$
limit that the widths (in radius, azimuthal angle, and polar
angle) of the spiral of radiation emitted by the rotating quark all go
to zero and the radiation becomes localized on the spiral (2.12).  For
$v<1$, the spiral of radiation is broadened and its location as a
one-dimensional curve is no longer uniquely defined. In the $v\rightarrow
1$ limit, however, the spiral of radiation and the spiral of the string
in the bulk are both one-dimensional curves, and there must be some
relation between these curves --- a relation that turns out to be simple
projection.

The shape of the rotating string determines the total power radiated, as we now explain.
The string has an energy density $\pi_0^0$ and energy flux $\pi_0^1$,
where
\begin{equation}
\pi_M^a \equiv  \frac{\delta S_{\rm NG}}
{\delta (\partial_a X^M)}
\end{equation}
with $S_{\rm NG}$ given in (\ref{SNG}), which satisfy
\begin{equation}
\partial_t \pi_{0}^{0} + \partial_u \pi_0^1 = 0\ .
\end{equation}
This is simply the statement of energy conservation
on the string worldsheet.
The energy flux down the rotating string is given by
\begin{equation}
\label{stringflux}
-\pi^1_0 = \frac{\sqrt{\lambda}}{2 \pi} \freq \Pi =\frac{\sqrt{\lambda}}{2\pi}\frac{1}{u_c^2}=
\frac{\sqrt{\lambda}}{2 \pi} a^2\, ,
\end{equation}
where we have used (\ref{PIandV}) and where $a = v\omega_0 \gamma^2$ is
the proper acceleration of the test quark, and where the minus sign
comes from our choice of parametrization of the world sheet with a coordinate $u$ that increases downward, away from the boundary. 
The
energy flux
down the string is the energy that must be supplied by the external
agent which moves the test quark in a circle; by energy conservation,
therefore, it is the same as the total power radiated
by the quark.\footnote
{  
The energy loss rate~\eqref{stringflux} can also be derived from the energy density along the string
     \begin{equation}
    -\pi_0^0 = {\sqrt{\lambda} \over 2 \pi} \, {\gamma \over u^2} \left(1+ \Pi \omega_0 u^2\right)\ ,
     \end{equation}
 as follows.
For $u\gg u_c$, this energy density approaches the constant
 \begin{equation}
-\pi_0^0 \approx  {\sqrt{\lambda} \over 2 \pi} \, \gamma \Pi \omega_0  \ .
 \end{equation}
Now, consider the total energy along a segment of the string whose extent in $u$ is $\delta u$, which we shall denote by $\delta E = - \pi_0^0 \delta u$.   The rotating quark can be thought of as ``producing this much new string'' in a time $\delta t$ that, from (\ref{varphiS}), is given by 
$\delta t= {\delta u \over \gamma}$. One should then identify $\delta E$ with the energy pumped into the system during $\delta t$.  Equating $\delta E = {d E \over dt} \delta t = P \delta t$ we thus find that
 \begin{equation}
 P =
  {1 \over \gamma} \pi_0^0  =  {\sqrt{\lambda} \over 2 \pi} \, \Pi \omega_0 \ .
 \end{equation}
}
Thus far, we have reproduced Mikhailov's result
(\ref{MikhailovPower}) via a calculation along the lines of Ref.~
\cite{Fadafan:2008bq} --- but with the advantage that we have been
able to find analytic expressions for $R(u)$ and $\phi(u)$, which will
be a significant help below.


We close this subsection by commenting on an aspect of the physics of the rotating string that we shall not pursue in detail.
As pointed out earlier, there is a special value of $u$, $u=u_c$ where we should specify the initial condition needed to solve the equations of motion that determine the shape of the rotating string.  
$u_c$ has a more physical interpretation: it is the depth at which the local velocity of the rotating string, $\freq R(u_c)$, becomes equal to the speed of light in the five-dimensional spacetime.  
This has a simple interpretation from the point of view of the string worldsheet. It can be verifed that the induced metric on the string worldsheet $g_{ab}$ has an event horizon at $u=u_c$. Thus disturbances on the string become causally disconnected across $u_c$. 
(This suggests that in the boundary 
field theory the regime $r \gg  R(u_c)=R_0/v$ can be thought of as the far field region,
while $r \lesssim R_0/v$ can be thought of as the near field region.)
Even though the AdS$_5$ spacetime with metric $G_{MN}$ has zero temperature, the worldsheet metric is that of a $(1+1)$-dimensional Schwarzschild black hole whose horizon is at $u=u_c$ and whose Hawking temperature is given by
     \begin{equation} \label{wste}
    T_{\rm ws} = {1 \over 2 \pi \gamma u_c} = {1 \over \gamma} T_{\rm Unruh}
    \end{equation}
where $T_{\rm Unruh} = {a \over 2 \pi}$ is the Unruh temperature for an accelerated particle with proper acceleration $a$. As it radiates, the rotating test quark
should experience small kicks which would lead to Brownian motion in coordinate space if the quark had finite mass.
At strong coupling such fluctuations can be found from small fluctuations of the worldsheet fields $X^M (t,u)$ in the worldsheet black hole geometry. The thermal nature of the worldsheet metric indicates that such Brownian motion can be described as if due to the presence of a thermal medium with a temperature given by~\eqref{wste}. 

\subsection{Gravitational perturbation set-up}

In the limit $\Nc \to \infty$, the $5d$ gravitational constant is
parametrically small and consequently the presence of the string
acts as a small perturbation on the geometry.  To obtain leading
order results in $1/\Nc$ we write the full metric as
\begin{equation}
G_{MN}= G^{(0)}_{MN} +h_{MN} \,,
\end{equation}
where $G^{(0)}_{MN}$ is the unperturbed metric (\ref{metric}), and
linearize the resulting Einstein equations in the perturbation $h_{MN}$.
This results in the linearized equation of motion
\begin{align}
    &- D^2 \, h_{MN} +2 D^{P} D_{(M} h_{N) P}
      -D_{M}D_{N} \, h +\coeff{8}{L^2} \, h_{MN}
\nonumber
\\ \label{lin1} &{}
    + \left (D^2 h -D^{P} D^{Q} \, h_{P Q} -\coeff{4 }{L^2} \, h \right)
G_{MN}^{(0)}  =  2 \kappa_5^2 \; t_{MN} \, ,
\end{align}
where $h \equiv h^{M}_{\ M}$, $D_M$ is the covariant derivative
under the background metric (\ref{metric}), $\kappa_5^2$ is the $5d$
gravitational constant and $t_{MN}$ is the $5d$ stress
tensor of the string.   In ${\cal N}=4$ SYM theory, $\kappa_5^2=4\pi^2 L^3/N_c^2$, but this relation would be different in other strongly coupled conformal field theories with dual gravitational descriptions. We shall see that $\kappa_5^2$ does not appear in any of our results.

According to the gauge/gravity dictionary, the on-shell gravitational action
\begin{equation}
S_G = \frac{1}{2\kappa_5^2}\int d^5 x \sqrt{-G}\left({\cal R}+\frac{12}{L^2}\right) + S_{GH}
\end{equation}
is the generating functional for the boundary stress
tensor~\cite{Witten:1998qj, deHaro:2000xn}.
Here, $G$ is the determinant of $G_{MN}$, ${\cal R}$ is its Ricci scalar, and $S_{GH}$ is the Gibbons-Hawking boundary
term~\cite{Gibbons:1976ue}, discussed and evaluated in the present context in Ref.~\cite{Chesler:2007sv}.
The $5d$ metric $G_{MN}$ induces a $4d$ metric $g_{\mu \nu}$ on the boundary of the the $5d$ geometry.
The boundary metric is related to the bulk metric by
\begin{equation}
g_{\mu \nu}(x) \equiv \lim_{u \to 0} \frac{u^2}{L^2} G_{\mu \nu}(x,u)\,.
\label{BoundaryMetric}
\end{equation}
Because $G_{MN}\propto 1/u^2$, the rescaling by $u^2$ in (\ref{BoundaryMetric}) yields a boundary metric that is regular at $u=0$.
The boundary stress tensor is then given by \cite{deHaro:2000xn}
\begin{equation}
\label{stress}
T^{\mu \nu}(x) = \frac{2}{\sqrt{-g}} \frac{\delta S_{\rm G}}{\delta g_{\mu \nu}(x)}\,,
\end{equation}
with $g$ denoting the determinant of $g_{\mu \nu}$.  We see that in order to
compute the boundary stress tensor, one needs to find the string profile dual to the rotating quark and
compute its $5d$ stress tensor $t_{MN}$.  One then solves the linearized Einstein equations (\ref{lin1}) in the
presence of the string source and then extracts the boundary stress
tensor from the variation of the on-shell gravitational action, as in (\ref{stress}).
The $\kappa_5^2$ dependence drops out because $S_G\propto 1/\kappa_5^2$ while we see from (\ref{lin1}) that the perturbation to the metric is proportional to $\kappa_5^2$.

We determined the string profile in Section III.A.   From this, we may now compute the $5d$ string stress
tensor, which is what we need in order to determine the metric
perturbation due to the string.  In general,
\begin{equation}
t^{MN}
=
-\frac{T_0}{\sqrt{-G}}\sqrt{-g}g^{ab}
\partial_a X^M \partial_b X^N
\delta^{3}(\bm r-\bm r_s)\,.
\end{equation}
For the rotating string given by (\ref{Xstring}) and (\ref{StringWorldsheetParametrization}) with (\ref{RSolution}) and (\ref{phiSolution}), this reduces to
\begin{widetext}
\begin{align}
\label{stringstress}
t_{MN} =  \frac{\sqrt{\lambda}}{2 \pi u L \sqrt{-g}} \delta^3(\bm r -
\bm r_s)
\left(
\begin{array}{ccccc}
R'^2+R^2  s^2 \phi '^2+1 & R^2 s^2 \freq  R' \phi ' & 0 & -R^2 s^2
\freq   (R'^2+1) & R^2 s^2\freq
    \phi ' \\ \\
R^2 s^2 \freq    R' \phi ' & R'^2 (R^2 s^2\freq ^2  -1) & 0 & -R^2
s^2  R' \phi '& R'
  \left(R^2 s^2 \freq ^2  -1\right) \\ \\
0 & 0 & 0 & 0 & 0 \\ \\
-R^2 s^2 \freq    (R'^2+1) & -R^2 s^2  R' \phi ' & 0 & R^4  s^4(\freq ^2
  R'^2+\freq^2 -\phi '^2) & -R^2  s^2 \phi ' \\ \\
R^2 s^2\freq    \phi ' & R' (R^2 s^2 \freq ^2 -1) & 0 & -R^2  s^2\phi
' & R^2 s^2 \freq ^2-1
\end{array}
\right) \, ,
\end{align}
\end{widetext}
where the components are in spherical coordinates, with the rows and
columns ordered $(t,r,\theta,\varphi,u)$, and where $s\equiv \sin\theta
$.

\subsection{Gauge invariants and the boundary energy density}

Given the string stress tensor (\ref{stringstress}), one can solve 
the linearized Einstein equations (\ref{lin1}).  However, 
doing this in its full glory is more work than is necessary.  
The metric perturbation $h_{MN}$ contains fifteen degrees of freedom
which couple to each other via the linearized Einstein equations.  
This should be contrasted with the $4d$ boundary stress tensor, which is traceless and 
conserved and thus contains five independent degrees of freedom.  Therefore,
not all of the degrees of freedom contained in $h_{MN}$ are physical.  
The linearized Einstein equations are invariant under infinitesimal coordinate transformations
$X^M \to X^M + \xi^M$ where $\xi^M$ is an arbitrary infinitesimal vector field.
Under such transformations, the metric perturbation transforms as
\begin{equation}
\label{redundancy}
h_{MN} \to h_{MN} - D_M \xi_N - D_N \xi_M\,,
\end{equation}
where $D_M$ is the covariant derivative under the background metric $G_{MN}^{(0)}$.
The physical information contained in $h_{MN}$ must be gauge invariant.  This requirement
restricts the number of physical degrees of freedom in $h_{MN}$ to five,%
\footnote
  {
  Five degrees of freedom in $h_{MN}$
  can be eliminated by exploiting the gauge freedom to set $h_{5M} = 0$ for all $M$,
  reducing the number of degrees of freedom in $h_{MN}$ to ten.  However, just as
  in electromagnetism, this does not completely fix the gauge.  There exists a residual 
  gauge freedom which allows one to eliminate five more components of $h_{MN}$
  on any $u = {\rm const.}$ slice. 
  }
matching that of the boundary stress tensor.

Just as in electromagnetism where gauge invariant quantities ({\em e.g.} electric and magnetic fields)
can be constructed from a gauge field, it is possible to construct gauge invariant quantities out
of linear combinations of $h_{MN}$ and its derivatives.  The utility of doing this lies in the fact that the
equations of motion for gauge invariants don't carry any of the superfluous gauge variant information
contained in the full linearized Einstein equations and thus can be simpler to work with.  
Indeed, as demonstrated in Refs.~\cite{Kovtun:2005ev, Chesler:2007sv, Chesler:2007an}, 
the equations of motion for gauge invariants can be completely decoupled from each other.

As the background geometry is translationally invariant, it is useful to introduce a $4d$ spacetime Fourier
transform and work with mode amplitudes $h_{MN}(u; \omega, \bm q)$.  Useful gauge invariants can 
then be constructed out of linear combinations of $h_{MN}(u; \omega, \bm q)$ and its radial
derivatives and classified according to their behavior under rotations about the ${\hat {\bm q}}$-axis.  
As $h_{MN}$ is a spin two field, there exists one independent helicity zero combination and a pair each of 
independent helicity one and two combinations.\footnote
  {
  There exist many different helicity zero, one and two gauge invariant 
  combinations of $h_{MN}$, but only five are independent.  
  Different gauge invariants of the same helicity are related to each other 
  by the linearized Einstein equations.
  } 

For utility in future work, we shall give an expression for the gauge invariant quantity that is valid at zero or nonzero temperature, even though in this paper we shall only use it at $T=0$.
At nonzero $T$,
the warp factor $f$ appearing in the metric (\ref{metric}) is 
given by (\ref{fBH}).  
When we later want to  recover $T=0$, we will simply set $f=1$.

We define 
$H_{MN} \equiv \frac{u^2}{L^2} h_{MN}$
and
\begin{align}
\nonumber
Z &\equiv  \frac{4 f}{\omega} q^i  H'_{0i} - \frac{4 f'}{\omega}q^i  H_{0i}  - \frac{2 f'  }{q^2} q^i q^jH_{ij} + 4 i f  q^i H_{i 5}
\\ \label{Zdef}
&- \frac{ (2 u q^2 {-} f')}{q^2} \left (q^2 \delta^{ij} {-}  q^i q^j  \right ) H_{ij}
+ \frac{4 q^2 f}{i \omega} H_{05} - \frac{8 \kappa_5^2 f}{i \omega} t_{05}\,,
\end{align}
where the zeroth and fifth coordinates are $t$ and $u$ respectively and where $i$ and $j$ run over the three spatial coordinates.
The quantity $Z$ clearly transforms as a scalar under rotations. With some effort, one can show that $Z$ is invariant under the infinitesimal gauge 
transformations (\ref{redundancy}).  $Z$ is therefore a suitable gauge invariant quantity for our purposes.

The dynamics of $Z$ are governed by the linearized Einstein
equations (\ref{lin1}).  Using the linearized Einstein equations, it is straightforward but tedious to show that $Z$ satisfies 
the equation of motion
\begin{equation}
\label{z0eqn}
Z'' + A Z' + B Z = S\,,
\end{equation}
where
\begin{eqnarray}
A &\equiv& -\frac{24 + 4 q^2 u^2 + 6 f + q^2 u^2 f - 30 f^2}{u f \left (u^2 q^2 + 6 - 6 f \right ) }\,,
\\
B &\equiv& \frac{\omega^2}{f^2} + \frac{  q^2 u^2(14 {-} 5 f {-} q^2 u^2) +18  (4{-}f{-} 3 f^2 )}{u^2 f \left ( q^2 u^2 +6 - 6f \right ) }\,,
\ \ \ \ \ \ \ \ 
\\ \label{generalsource}
\frac{S}{\kappa_5^2} &\equiv& \frac{8 }{f} t'_{00}  
+ \frac{4 \left ( q^2 u^2 {+}6 {-} 6f \right ) }{3 u q^2 f} (q^2 \delta^{ij} {-} 3 q^i q^j ) t_{ij}
\\ \nonumber
&+&  \frac{8 i \omega}{f} t_{05}
+ \frac{8u \left[   q^2  \left(q^2 u^2{+}6\right) - f \left ( 12 q^2  {-} 9 f'' \right )\right]  }{3  f^2 \left(q^2 u^2-6 f+6\right)} t_{00}
\\ \nonumber
&-&\frac{8 q^2 u}{3} t_{55}
- 8 i q^i t_{i5}\,.
\end{eqnarray}

The connection
between $Z$ and the energy density may be found by
considering the behavior of $Z$ and $H_{MN}$ near the boundary.  
Choosing for convenience the gauge $H_{5M} = 0$, one can solve the linearized 
Einstein equations (\ref{lin1}) with a power series expansion about $u = 0$ in order to ascertain
the asymptotic behavior of the metric perturbation.
Setting the boundary value of $H_{\mu \nu}$ to vanish so the boundary geometry is flat
and considering sources $t_{MN}$ corresponding to strings ending at $u = 0$, one
finds an expansion of the form \cite{Friess:2006fk, Chesler:2007sv}
\begin{equation}
\label{Hexpansion}
H_{\mu \nu}(u)=  H_{\mu \nu}^{(3)} \, u^3 + H_{\mu \nu}^{(4)} \, u^4
    + \cdots.
\end{equation}
In the gauge $H_{5M} = 0$ the variation of the 
gravitational action (\ref{stress}) relates the asymptotic behavior of $H_{\mu \nu}$
to the perturbation in the boundary energy density via \cite{deHaro:2000xn}
\begin{equation}
\mathcal E = \frac{2 L^3}{\kappa_5^2} H_{00}^{(4)}\,.
\end{equation}
The coefficient $H_{00}^{(4)}$ can in turn be related to the asymptotic 
behavior of $Z$ by substituting the expansion (\ref{Hexpansion}) into Eq.~(\ref{Zdef}).
In doing so one finds that $Z$ has the asymptotic form
\begin{equation}
\label{zexpan}
Z(u) =   Z_{(2)} \, u^2 +   Z_{(3)} \, u^3+ \cdots,
\end{equation}
and that
\begin{eqnarray}
H_{00}^{(4)} = -\frac{1}{16} Z_{(3)}\,.
\end{eqnarray}
We therefore see that the energy density is given by
\begin{equation}
\label{eq:EfromZ3}
\mathcal E = -\frac{ L^3}{8 \kappa_5^2} Z_{(3)}\,.
\end{equation}
The coefficient $Z_{(2)}$ in the expansion (\ref{zexpan}), which has delta function support at the location of the quark, gives the divergent stress of the infinitely massive test quark. It is therefore not of interest to us.
We now see explicitly that, as we argued above, in order to obtain the energy density in the boundary quantum field theory the only aspect of the metric perturbation that we need to compute is $Z$, and furthermore that all we need to know are the coefficients in the expansion of $Z$ about $u=0$.



\subsection{The solution to the bulk to boundary problem and the boundary energy density}


Although we have defined $Z$ at nonzero temperature, henceforth as we determine $Z$ we return to $T=0$, meaning $f=1$.
At zero temperature the coefficients $A$ and $B$ appearing in (\ref{z0eqn}) 
are given by
\begin{eqnarray}
A &=& -\frac{5}{u}\,,
\\
B &=& \omega^2 - q^2 + \frac{9}{u^2}.
\end{eqnarray}
and the general solution to (\ref{z0eqn}) may be written
\begin{align}
\nonumber
Z(u) = &-u^3 I_0(u Q) \left [ \int_u^\infty du' \frac{ K_0(u' Q)}{ u'^2} S(u') + \alpha \right ]
\\ \ \ \ \ \  \label{solution}
&-u^3 K_0(u Q) \lim_{\epsilon \to 0} \left [\int_\epsilon^u du' \frac{I_0(u' Q)}{u'^2} S(u') + \beta \right ]\,,
\end{align}
where $Q \equiv \sqrt{q^2 - \omega^2}$, 
$I_0$ and $K_0$ are modified Bessel functions,
and $\alpha$ and $\beta$ are constants of integration.

The constants of integration are fixed by requiring that $Z(u)$ satisfy appropriate boundary conditions.
As $I_0( u Q)$ diverges as $u \to \infty$, regularity at $u = \infty$ requires $\alpha  = 0$.  The constant
$\beta$ is fixed by the requirement that the $\epsilon \to 0$ limit exists (so all points 
on the string contribute to the induced gravitational disturbance) and that no logarithms
appear in the expansion of $Z(u)$ near $u = 0$.  This last condition is equivalent to the boundary
condition that the metric perturbation $H_{MN}$ vanish at $u = 0$ so the boundary geometry is flat
and unperturbed.
For strings which end at $u = 0$, we have 
\begin{equation}
\label{sexp}
S(u) = s_0 + \mathcal O(u^2)\,,
\end{equation}
where
\begin{equation}
s_0 =8 \kappa_5^2  \lim_{u \to 0}  u^2 \partial_u\left( \frac{t_{00}}{u^2}\right) = -8\kappa_5^2 \lim_{u \to 0} \frac{t_{00}}{u}\,,
\end{equation}
where we have used the fact that $t_{00}\propto u$ at small $u$.
Furthermore, the Bessel functions have the asymptotic expansions
\begin{align}
\label{I0exp}
I_0(u Q) &= 1 + \mathcal O(u^2)\,,
\\ \label{k0exp}
K_0(u Q) &= -\gamma_{\rm E}  - \log ( {\textstyle \frac{1}{2}} u Q) + \mathcal O(u^2 )\,,
\end{align}
where $\gamma_{\rm E}$ is Euler's constant.  
Substituting these expansions into the second integral in (\ref{solution}), we find
\begin{equation}
\int_\epsilon^\infty du' \frac{ I_0(u' Q)}{ u'^2} S(u')  =  \frac{s_0}{\epsilon} + \mathcal O(1)\,.
\end{equation}
Therefore, in order for the $\epsilon \to 0$ limit to exist we must have 
\begin{equation}
\label{beta}
\beta = -\frac{s_0}{\epsilon}\,.
\end{equation}
As one may readily verify, this value of $\beta$ also eliminates all logarithms appearing in the expansion of 
$Z$ near the boundary.%
\footnote
  {
  A deformation in the boundary geometry implies that 
  logarithms will occur in the expansion of $Z$ at order $u^3$ and beyond.
  The linearized Einstein equations relate all higher order log coefficients
  to that of the $u^3$ coefficient in a linear manner.
  Therefore, if the order $u^3$ coefficient vanishes all other coefficients vanish.
  As is easily seen from the asymptotic forms (\ref{sexp}) -- (\ref{k0exp}), the value of $\beta$
  given in (\ref{beta}) ensures the order $u^3$ logarithm coefficient vanishes.
  }

With constants of integration determined and with the asymptotic forms (\ref{sexp})--(\ref{k0exp}), 
it is easy to read off from (\ref{solution}) the asymptotic behavior of $Z$. In particular, we find
\begin{equation}
\label{Z3}
Z_{(3)} =  \lim_{\epsilon \to 0} \left  [ \int_ \epsilon ^\infty \!\! du \,\mathcal G(u)  S(u)  - s_0 \left ( \frac{1}{ \epsilon}  + \epsilon \, \mathcal G(\epsilon) \right )\right ]\,,
\end{equation}
where 
\begin{equation}
\mathcal G \equiv -\frac{K_0(u \sqrt{q^2 - \freq^2})}{ u^2}\,,
\end{equation}
is the \textit{bulk to boundary} propagator.

At this point it is convenient to Fourier transform back to real space.  At zero temperature
the source (\ref{generalsource}) Fourier transforms to
\begin{align}
\frac{S(t,\x,u)}{\kappa_5^2}
&=
8u^2 \partial_u \left(\frac{t_{00}}{u^2}\right)
- \frac{4 u }{3}\nabla^2 (2t_{00}-2t_{55}+ t_{ii})
\nonumber\\
&
+ 4 u  \nabla_i \nabla_j t_{ij}
- 8 \partial_t t_{05}
- 8 \nabla_i t_{i5}
\,,\label{SFourier}
\end{align}
where sums over repeated spatial indices $i$ and $j$ are implied.
To take the Fourier transform of the bulk to boundary propagator
one must give a prescription for integrating around singularities at $\omega = q$.
The relevant prescription comes from causality and requires analyticity in
the upper half frequency plane.  This requires sending $\omega \to \omega + i \epsilon$.
With this prescription we have
\begin{equation}
\mathcal{G}(t,\bm r,u)
=
\frac{1}{\pi u^2}\theta(t)\delta'(-t^2+\bm r^2+u^2)\,.
\end{equation}
We therefore have
\begin{align}
\nonumber
Z_{(3)}(t,\bm r) =& \lim_{\epsilon \to 0} \int \! d^4 r'  \bigg [  \int_{\epsilon}^{\infty} \!\!\!\! du \,
\mathcal{G}(t{-}t',\bm r{-}\bm r',u) S(t',\bm r',u) 
\\ \label{Z3real}&- \epsilon \, s_0(t',\bm r') \mathcal G(t',\bm r {-} \bm r', \epsilon) \bigg ]
- \frac{s_0(t,\bm r)}{\epsilon}\,.
\end{align}
When we  use (\ref{SFourier}) to express $S(t',\bm r',u)$ we find that when the $du$-integral of the first term in (\ref{SFourier}) is done by parts, the resulting  boundary term cancels the $\epsilon$-dependent bulk-to-boundary propagator  in (\ref{Z3real}).


Upon assuming that the stress-energy tensor on the boundary does not have support at $(t,\r)$, i.e. that the observer is located away from the source, we can safely take the $\epsilon \to 0$ limit and  obtain for the energy density
\begin{eqnarray}
\mathcal{E}(t,\r) &=&
\frac{L^3}{\pi}
\int d^4r'
\int_0^\infty \!\! du \,\,
\theta(t-t')
\nonumber\\&&
\Big[
\left(
4u t_{00}
-
t_{M5}\nabla'_M \mathcal{W}
\right)
\frac{\delta''(\mathcal{W})}{u^2}
\nonumber\\
&&
+\vert \r-\r' \vert^2
\left(
4t_{00}-4t_{55}+2t_{ii}
\right)
\frac{\delta'''(\mathcal{W})}{3u}
\nonumber\\
&&
-
\left(
t_{ij}\nabla'_i \mathcal{W} \nabla'_j \mathcal{W}
\right)
\frac{\delta'''(\mathcal{W})}{2u}
\Big]
\,,
\end{eqnarray}
where
\begin{eqnarray}
\mathcal W  &\equiv&  -(t-t')^2 + u^2 +|\r-\r'|^2
\end{eqnarray}
and $\nabla'_M$ is the partial derivative with respect to $X'^M=\{t',\r',u\}$.  We see that the $\kappa_5^2$ factors in (\ref{eq:EfromZ3}) and (\ref{SFourier}) have cancelled, as expected.
Because we have analytic expressions for the profile of the rotating string, it turns out that we can evaluate ${\cal E}$ explicitly, in two stages.  First, using (\ref{stringstress}) including its three-dimensional delta function, and using our expressions (\ref{RSolution}) and (\ref{phiSolution}) for $R(u)$ and $\phi(u)$, we obtain 
\begin{equation}
\label{eq:EE2E3}
\mathcal{E}(t,\x) =
\frac{\sqrt{\lambda}}{\pi^2}\int_0^\infty\!\!\! du \!\!
\int_{-\infty}^t \!\!\!dt' \left [\mathcal{A} \, \delta''(\mathcal W_s) + \mathcal{B}\, \delta'''(\mathcal W_s) \right ]
\,,
\end{equation}
with
\begin{eqnarray}
\mathcal{A}
&=&
\gamma  \left(3 +  v^2 \gamma ^3 \omega _0^2 u \left(-t'+t+u \gamma\right)\right)
-
v \gamma  \omega _0 r\sin\theta\cos\psi
\nonumber\\&&
+  v^3 \gamma ^4 \omega _0^2 u r\sin\theta\sin\psi 
\,,
\nonumber\\
\mathcal{B}
&=&
v^4 \gamma ^3 u^2 \left(1- v^2 \gamma ^4 \omega _0^2u^2\right)
\nonumber\\&&
+
\left( v^4 \gamma ^5 \omega _0^2 u^2+\gamma +\frac{1}{3 \gamma }\right)|\r-\r_s(t',u)|^2
\nonumber\\&&
-
2 v^3 \gamma ^3\omega _0 u^2 \left(\cos\psi- v^2 \gamma ^3 \omega _0 u \sin\psi\right)
r \sin\theta
\nonumber\\&&
- v^2 \gamma
(1 + v^2 \gamma ^4 \omega _0^2 u^2) r^2\sin^2\psi\,  \sin^2\theta
\nonumber\\&&
+ 2 v^2 \gamma^2 \omega _0 u\, r^2\cos\psi \,\sin\psi\,  \sin^2\theta
\,,
\end{eqnarray}
where $\psi=\varphi + \omega_0 (\gamma u - t')$ and
\begin{eqnarray}
\mathcal{W}_s \equiv -(t-t')^2 + u^2 +|\r-\r_s(t',u)|^2
\,.
\end{eqnarray}
Next, the integral (\ref{eq:EE2E3}) can be carried out via the change of variables
\begin{equation}
\zeta= \frac{1}{2}(\gamma u+t'-t)  \,,\quad \nu=\frac{1}{2}(\gamma u-t'+t)
\,.
\end{equation}
Since points with $-(t'-t) < \gamma u$ are causally disconnected from an observer on the boundary at time $t$, the argument of the $\delta$-functions in Eq.~(\ref{eq:EE2E3}) is non-vanishing.
This allows us to deform the domain of integration and the integral takes the generic form
\begin{eqnarray}
I_n
&=&
\int_0^\infty du \int_{-\infty}^{t-\gamma u}dt' F_n(t',u) \delta^{(n)}(\mathcal{W}_s)
\,
\nonumber\\
&=&
\frac{2}{\gamma}\int_0^\infty d\nu \int_{-\nu}^{0}d\zeta \tilde{F}_n(\zeta,\nu)  \delta^{(n)}(\mathcal{W}_s)
\,,
\end{eqnarray}
where $\tilde{F}_n(\zeta,\nu)=F_n(t'(\zeta,\nu),u(\zeta,\nu))$.
Furthermore, the argument of the $\delta$-functions takes the form
\begin{eqnarray}
\mathcal{W}_s
&=&
r^2+R_0^2
+
\zeta \left(4\nu - 2  v r \sin\theta  \sin (2 \nu \omega_{0} +\varphi - \omega_0 t )\right)
\nonumber\\&&-
2 r R_0 \sin \theta [\nu \omega_{0}  \sin (2 \nu \omega_{0} +\varphi - \omega_0 t )
\nonumber\\&&\hspace{2.5cm}
+ \cos (2 \nu \omega_{0} +\varphi -\omega_0 t)]
\,,
\end{eqnarray}
which is then linear in $\zeta$ and therefore allows us to evaluate $I_n$ formally, obtaining
\begin{eqnarray}
\label{eq:In}
I_n
&=&
\frac{2}{\gamma}
\int_{\nu_0}^\infty d\nu
\,\,\,
\left.
\frac{
(-1)^n\partial_\zeta^n \tilde{F}_n(\zeta,\nu)
}{
(\partial_\zeta \mathcal{W})^{n+1}
}
\right|_{\zeta=\zeta_0(v)}
\nonumber\\&+&
\frac{2}{\gamma}
\int_0^\infty d\nu
\sum_{l=1}^n
\frac{
(-1)^{l}
}{
(\partial_\zeta \mathcal{W})^l
}
\left.\partial_{\zeta}^{l-1}\tilde{F}_n(\zeta,\nu)\right|_{\zeta=-\nu}
\nonumber\\&&
\phantom{aaa}\times
\delta^{(n-l)}(\mathcal{W}_s(t'=t-2\nu,u=0))
\,.
\end{eqnarray}
Here $\zeta_0(v)$ is the unique zero of $\mathcal{W}_s$ at a given $\nu$ and $\nu_0=(t-t_{\rm ret})/2$, but we will not describe in detail how these arise because the integrand in the first term in $I_n$ vanishes:
by direct inspection of $F_2\equiv\mathcal{A}$ and $F_3\equiv\mathcal{B}$ we find
$\partial_\nu^2 \tilde{F}_2(\zeta,\nu)=0$
and
$\partial_\nu^3 \tilde{F}_3(\zeta,\nu)=0$.
Consequently only the boundary terms in the second line of Eq.~(\ref{eq:In}) contribute and we can perform
the remaining integral using the $\delta$-function.  
Finally, we obtain
\begin{eqnarray}
\label{eq:Eanalytical}
\nonumber
\mathcal E &=& \frac{\sqrt{\lambda}}{24 \pi^2 \gamma^4 r^6 \Xi^6}
\Bigg [
-2 r^2 \Xi ^2+4 r \gamma^2 \Xi  (t_{\rm ret}{-}t) \\ \nonumber
& &\quad+(2 \gamma^2 {-} 4 r^2 v^2 \gamma^2 \freq^2 \sin^2\theta{+}3 r^2 \gamma^4\freq^2\Xi^2)(t_{\rm ret}{-}t)^2 {\phantom\int}
\\ \nonumber
& & \quad + 7 r \gamma^2\freq^2\Xi (t_{\rm ret}{-}t)^3   + 4 \gamma^2\freq^2 (t_{\rm ret}{-}t)^4{\phantom\int}
\\ \nonumber
&  & \quad +8v\gamma^2\freq r   (t_{\rm ret}{-}t)   (t_{\rm ret}{-}t { + }r \, \Xi )
\\ 
& & \qquad\qquad\qquad\qquad\quad \times \sin \theta   \cos (\varphi {-} \freq t_{\rm ret}  )
\Bigg ]\,,\label{eq:Ex}
\end{eqnarray}
where $\Xi$ was given in (\ref{xieq}).
%
Recalling the discussion after (\ref{tret0}), we note that $\mathcal E$ is invariant under a shift  of $\varphi$ by $\delta\varphi$ that is compensated by shifts in both $t$ and $t_{\rm ret}$ by $\delta\varphi/\freq$.  
$\mathcal E$ is a periodic function of $\varphi$ with period $2\pi$ and a periodic function of  $t$ with period $2\pi/\freq$, as it must be.

%

Eq.~(\ref{eq:Ex}) is the main result of our paper.  It is an explicit analytic expression for the energy density of the radiation emitted by a test quark in circular motion, valid at all times and at all distances from the quark.  The only aspect of the calculation that requires numerical evaluation is that one must solve the transcendental equation (\ref{tret0}) for $t_{\rm ret}$.  With $t_{\rm ret}$ in hand, one uses (\ref{xieq}) to compute $\Xi$ and then evaluates the energy density $\mathcal E$ using the solution (\ref{eq:Ex}).  We shall illustrate our result (\ref{eq:Ex}) in several ways in Section IV.

\subsection{Far zone and angular distribution of power}

We close this section by evaluating the energy density (\ref{eq:Ex}) in the $r\to\infty$ limit
and extracting the angular distribution of power \`a la Eq.~(\ref{power1}).  In the far zone 
limit we can replace $\Xi$ by its $r\to\infty$ limit, namely $\xi$
in (\ref{xiInftyDefn}), and we can safely replace $t_{\rm ret}$ by $(t-r)$ everywhere in (\ref{eq:Ex}) except within the $\varphi$-dependent argument of the cosine.
In the far zone, the energy density (\ref{eq:Ex})  then reduces to that of Eq.~(\ref{farzonelimits}) with 
\begin{equation}
\label{farzoneeps}
\varepsilon =  
\frac{\freq^2 \sqrt{\lambda }}{24 \pi ^2 }
\frac{4-4v^2 \sin^2 \theta - 7 \xi + 3 \xi^2 \gamma^2}{\xi^6 \gamma^2} \,.
\end{equation}
Using Eq.~(\ref{power1}), we find that the time-averaged angular distribution
of power is given by
\begin{align}
\label{dpdOmega}
\frac{dP}{d \Omega}
= \frac{v^2 \freq^2 \sqrt{\lambda}}{16 \pi^2}
\frac{ 
	2 + v^2 \sin^2 \theta
}{
  \left(1- v^2 \sin ^2\theta \right)^{5/2}
}\ ,
\end{align}
which is the same as the weak-coupling result (\ref{VectorPlusScalar}) up to a $\theta$-independent overall factor!

Upon integrating over all solid angles, we find the total power radiated
\begin{equation}
\label{totalpower}
P = \frac{\sqrt{\lambda}}{2 \pi} a^2\,,
\end{equation}
where again $a=v\gamma^2\omega_0$ is the quark's proper acceleration, reproducing Mikhailov's result (\ref{MikhailovPower})  again. 
The fact that the power that we have obtained in this section by integrating over the 
angular distribution of the radiation (\ref{dpdOmega}) in the quantum field theory matches
the energy flux (\ref{stringflux}) flowing down the classical string in the dual gravitational description is a nontrivial check of our calculations.

\section{Results and Discussion}
\label{discussion}

\begin{figure*}[t]
\includegraphics[scale=0.38]{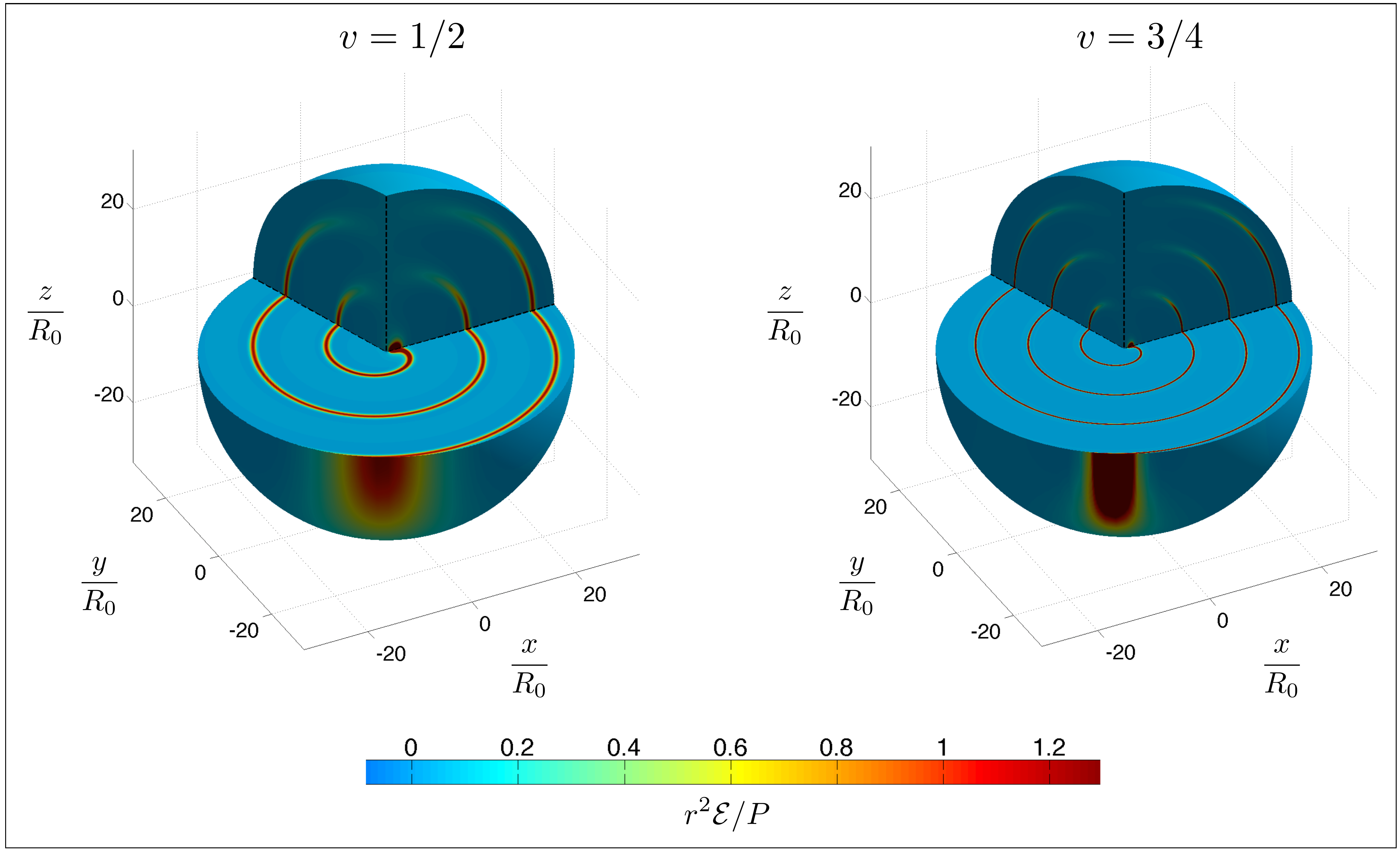}
\caption
    {\label{blingbling}
   Left: a cutaway plot of $r^2 \mathcal E/P$ for $v = 1/2$.
   Right: a cutaway plot of $r^2 \mathcal E/P$ for $v = 3/4$.
   In both plots the quark is at $x = R_0$, $y=0$ at the time shown and its 
   trajectory lies in the plane $z = 0$.  The cutaways coincide with the planes 
   $z = 0$, $\varphi = 0$ and $\varphi =  7 \pi / 5$.  At both velocities
   the energy radiated by the quark is concentrated along a spiral structure
   which propagates radially outwards at the speed of light.  The spiral is 
   localized about $\theta = \pi/2$ with a characteristic width $\delta \theta \sim 1/\gamma$.
   As $v \to 1$ the radial thickness $\Delta$ 
   of the spirals rapidly decreases like $\Delta \sim 1/\gamma^3$.  
   }
\end{figure*}

\begin{figure*}[t]
\includegraphics[scale=0.54]{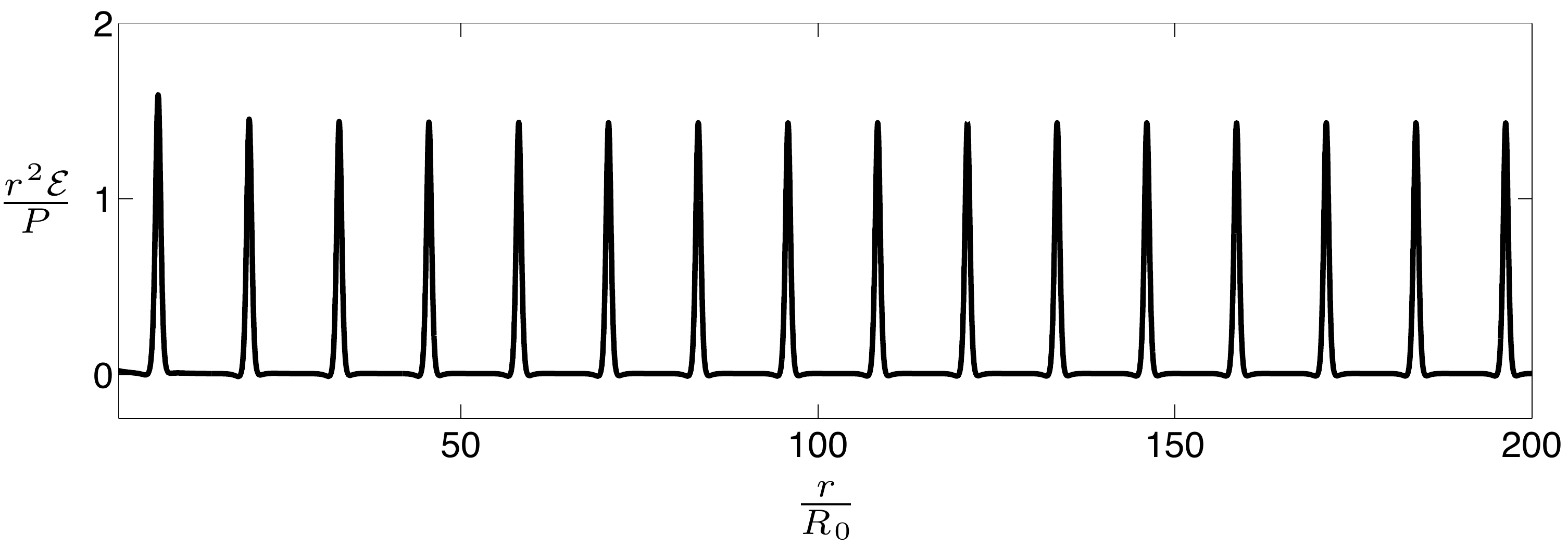}
\caption
    {\label{ekgplot}
 Plot of $r^2 \mathcal E/P$ at $\theta=\pi/2$ and $\varphi=5\pi/4$ at $t=0$ as a function of $r$ for $v=1/2$. The plot illustrates the fact that the pulses of radiated energy do not broaden as they propagate outward.  This implies that they do not broaden in azimuthal angle, either.  Strongly coupled synchrotron radiation does not isotropize.
    }
\end{figure*}

\subsection{Radiation at strong coupling, illustrated}

Fig.~\ref{blingbling} shows two cutaway plots of the energy density $\mathcal E$
at strong coupling (multiplied by $r^2/P$ where 
$P$ is the total power radiated) produced by a quark moving 
on a circle of radius $R_0$ at velocities $v = 1/2$ 
and $v = 3/4$.  The figure is obtained by evaluating (\ref{eq:Ex}).
The motion of the quark is confined to the plane $z = 0$ and 
at the time shown the quark is at $x = R_0$, $y = 0$, and is rotating 
counter clockwise. The cutaways in the plots show the energy density 
on the planes $z = 0$, $\varphi = 0$ and $\varphi = 7 \pi / 5$ where $\varphi$ is the azimuthal angle.   
As is evident from the figure, as the quark accelerates along its trajectory energy is radiated outwards in a spiral 
pattern.  This radiation falls off like $1/r^2$ and hence has a constant
amplitude in the figure and propagates radially outwards at the speed of light.   
The figure shows the location of the energy density at one time; as a function of time, the entire pattern of energy density rotates with constant angular frequency $\omega_0=R_0 v$. This rotation of the pattern is equivalent to propagation of the radiation outwards at the speed of light.
As seen by an observer far away from the quark, the radiation appears as a short pulse 
just like a rotating lighthouse beam does to a ship at sea.

What we see in  Fig.~\ref{blingbling} looks like an outward going pulse of radiation that does not broaden as it propagates.   Analysis of (\ref{eq:Ex}) confirms this, as we have illustrated in Fig.~\ref{ekgplot} by extending the plot of the energy density outward to much larger radii
at one value of the angular coordinates, at time $t=0$.  As time progresses, the pulses move outward at the speed of light, and an observer at large $r$ sees repeating flashes of radiation.  This figure provides a convincing answer to a central question that we set out to answer in Section I.  We see that even though the gauge theory is nonabelian and strongly coupled, the narrow pulses of radiation propagate outward without any hint of broadening.   We find the same result at much larger values of $\gamma$, where the pulses become even narrower --- their widths are proportional to $1/\gamma^3$, as at weak coupling and as we shall discuss below.   The behavior that we find is familiar from classical electrodynamics --- in which the radiated energy is carried by noninteracting photons.  Here, though, the energy is carried by fields that are strongly coupled to each other.  And, yet, there is no sign of this strong coupling in the propagation of the radiation.  No broadening. No isotropization.

Fig.~\ref{ekgplot} is also of some interest from a holographic point of view.   The qualitative idea behind gauge/gravity duality is that depth in the 5th dimension in the dual gravitational description corresponds to length-scale in the quantum field theory.   In many contexts, if one compares two classical strings in the gravitational description which lie at different depths in the 5th dimension, the string which is closer to the boundary corresponds to a thinner tube of energy density in the quantum field theory while the string which is deeper, farther from the boundary, corresponds to a fatter tube of energy density.    Our calculation shows that this intuitive way of thinking about gauge/gravity duality need not apply.  The rotating string falls deeper and deeper into the 5th dimension with each turn of its coils and yet the thickness of the spiral tube of energy density in the quantum field theory that this string describes changes not at all.   

The behavior of the  outgoing pulse of radiation illustrated in Fig.~\ref{ekgplot} is different than what one may have expected for a nonabelian gauge theory given that solutions to the {\it classical} field equations in $SU(2)$  and $SU(3)$ gauge theory are chaotic with positive Lyapunov exponents and are thought to be ergodic~\cite{Muller:1992iw,Biro:1993qc,Bolte:1999th}.  Of course, we have not solved classical field equations; we have done a fully quantum mechanical analysis of the radiation in a nonabelian gauge theory, in the limit of large $N_c$ and strong coupling. It is nevertheless surprising from the gauge theory perspective that the radiation we find turns out to behave (almost) like that in classical electrodynamics.  From the gravitational perspective, we saw that the equations describing the five-dimensional metric perturbations linearize, meaning that there is no possibility of chaotic or ergodic dynamics.

\begin{figure}[t]
\includegraphics[scale=0.51]{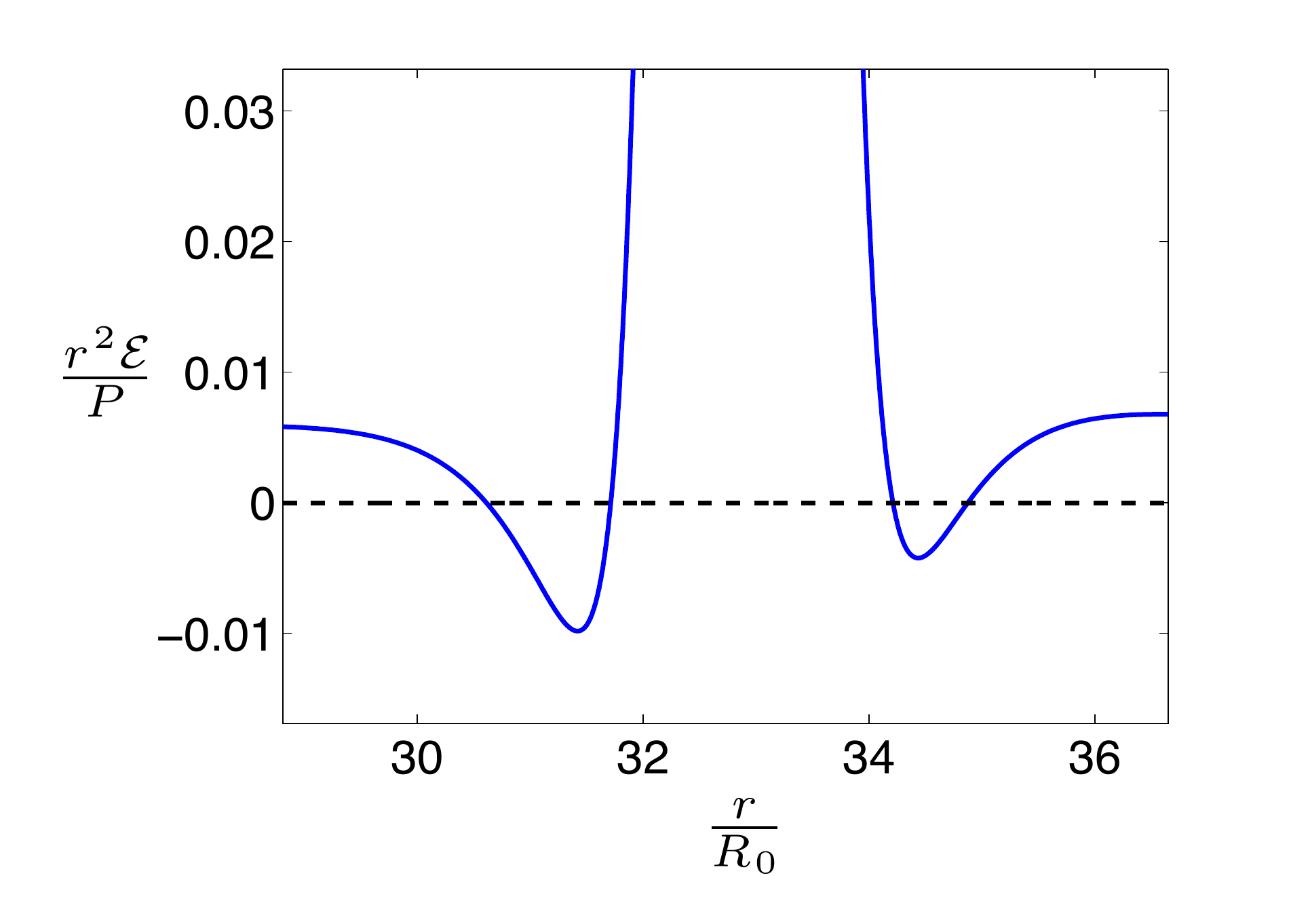}
\caption
    {\label{negativeenergy}
   The energy density at $\theta = \pi/2$ and $\varphi = 5 \pi/4$ for 
   quark velocity $v = 1/2$, as in Fig.~\protect{\ref{ekgplot}}.   $r \approx 33 R_0$ corresponds to the 
   location of a spiral in the energy density.  Directly ahead of and directly behind the spiral, the energy
   density is slightly negative.  To see how slightly, compare the vertical scale here with that in Fig.~\protect{\ref{ekgplot}}.
   }
\end{figure}

The synchrotron radiation that we have found at strong coupling is, in at least one respect, qualitatively different from that in familiar classical electrodynamics.  The difference is in principle visible in Fig.~\ref{ekgplot}, but the effect is small and hard to see without zooming in --- which we do in Fig.~\ref{negativeenergy}.  For large enough $v$ we find that
the energy density is {\em negative} in some regions of space!  
Directly ahead of and behind the spiral, 
the energy density dips slightly below zero.  This is impossible in classical electrodynamics.  The effect that we have found is numerically small, as a comparison of the vertical scales in Figs.~\ref{negativeenergy} and \ref{ekgplot} makes clear, but it is nevertheless in stark contrast to the weak coupling results of Section~\ref{weakcoupling}, where the energy density is always positive, as in classical electrodynamics.  This small effect serves as a reminder that the calculation that we have done is quantum mechanical. 
In a quantum field theory 
the energy density need not be positive everywhere --- only its integral over all space is constrained to be positive.
In a strongly coupled quantum field theory, quantum effects should be large.  So, it is not surprising that we see a quantum mechanical effect.  What is surprising is that it is so small, and that in other respects Figs.~\ref{blingbling} and \ref{ekgplot} look so similar to synchrotron radiation in classical electrodynamics.

\subsection{Synchrotron radiation at strong and weak coupling}

Given the qualitative similarities between Figs.~\ref{blingbling} and \ref{ekgplot} and the physics of synchrotron radiation in classical electrodynamics and weakly coupled ${\cal N}=4$ SYM theory, we shall attempt several more quantitative comparisons.  
First, as can be seen from Fig.~\ref{blingbling}, the radiation emitted by the quark
is localized in polar angle about the equator, $\theta = \pi/2$.  From the far-zone expression for the 
energy density, Eqs.~(\ref{farzonelimits}) and (\ref{farzoneeps}), it is easy to see that the characteristic
opening angle about $\theta= \pi/2$ scales like $\delta \theta \sim 1/\gamma$ in the $v \to 1$ limit.
Furthermore, from Fig.~\ref{blingbling} it can be seen that the radial thickness $\Delta$ of each pulse 
of radiation decreases as $v$ increases.  
Again, from Eqs.~(\ref{farzonelimits}) and (\ref{farzoneeps}) it is easy to see that 
$\Delta \sim 1/\gamma^3$ in the $v \to 1$ limit.\footnote
{
There may be a holographic interpretation of
$\Delta\sim 1/\gamma^3$.  We saw in Section~\ref{RotatingStringSection} that there is one special point on the string, namely the worldsheet horizon at $u=u_c$.  Using the correspondence between $u$ and length-scale in the boundary quantum field theory, we expect $u_c$ to translate into a length scale in the rest frame of the rotating quark, corresponding to a length scale
\begin{equation}
\label{delid}
 \delta \sim {u_c \over \gamma} = {1 \over \gamma^3 v \omega_0} = {R_0 \over \gamma^3 v^2} \ ,
  \end{equation}
  in the inertial frame in which the center of motion is at rest. It is tempting to 
  identify $\delta$ with $\Delta$ at $v\to 1$.
}  
Intriguingly, the scaling
of both $\delta \theta$ and $\Delta$ with $\gamma$ are identical to those at weak coupling.

\begin{figure}[t]
\includegraphics[scale=0.54]{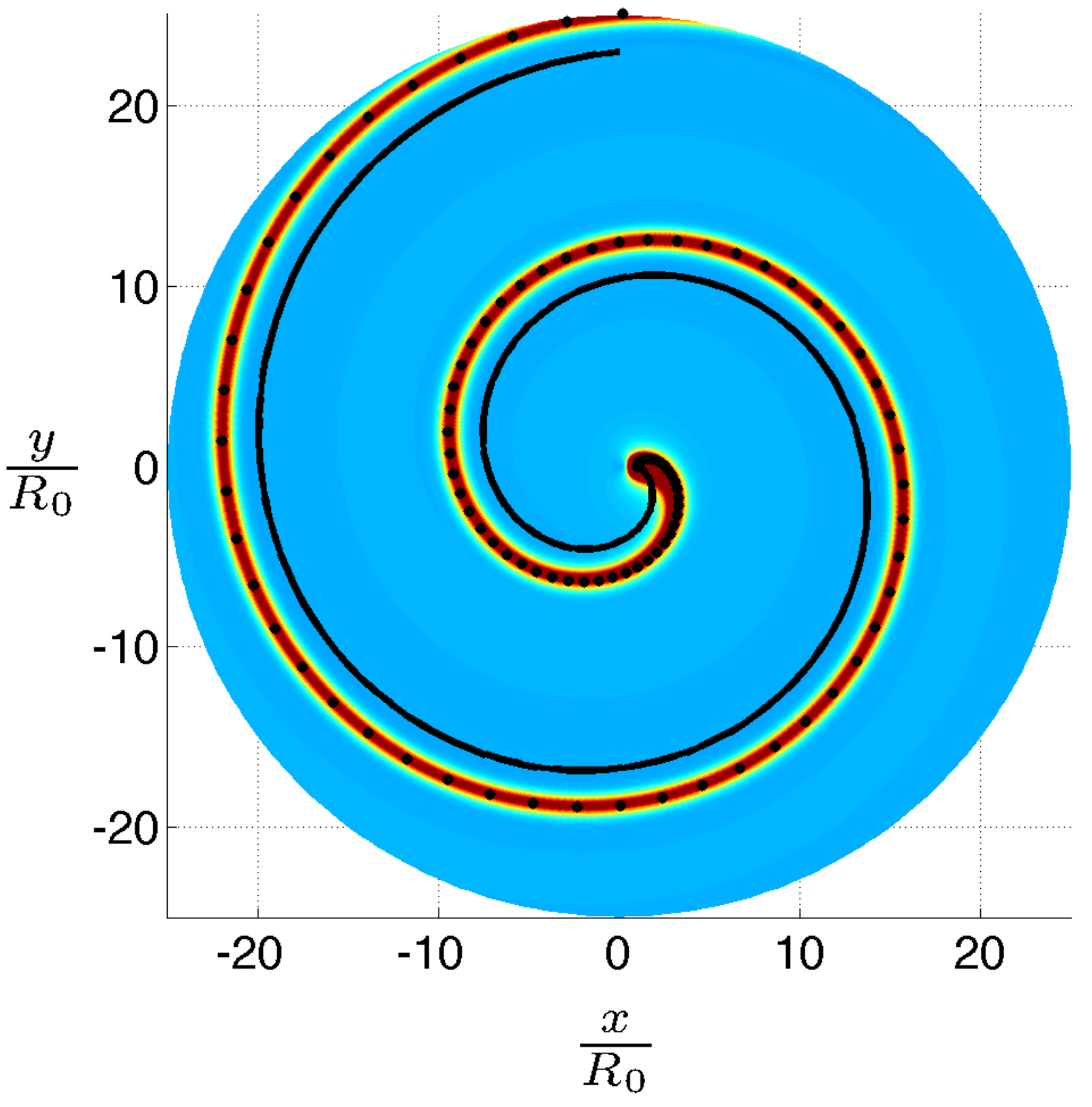}
\caption
    {\label{spiraldiagram}
$r^2{\cal E}/P$ for $v=1/2$ in the $z=0$ plane.    The color code is the same as in Fig.~{\protect \ref{blingbling}}, with zero energy density blue and maximal energy density red. The spiral curve marked with the black dots is ({\protect \ref{curve}}), namely the place where the spiral of synchrotron radiation would be in electrodynamics or in weakly coupled ${\cal N}=4$ SYM theory. We see that the spiral of radiation in the strongly coupled gauged theory with a gravitational dual is at the same location.  This indicates that, as at weak coupling, strongly coupled synchrotron radiation is beamed in the direction of the motion of the quark. For reference, the solid black line is ({\protect \ref{WrongSpiral}}), namely the place where the synchrotron radiation would be  if the quark were emitting a beam of radiation perpendicular to its direction of motion.  
    }
\end{figure}

Emboldened by the agreement between the $\gamma$-scaling of $\Delta$ and $\delta\theta$ at weak and strong coupling, we have compared the shape of the spiral of energy density in the $z=0$ plane, for example as depicted at two velocities in Fig.~\ref{blingbling}, with the shape of the classic synchrotron radiation spiral (\ref{curve}).  We find precise agreement, as illustrated in Fig.~\ref{spiraldiagram}. At strong coupling, the energy density $\cal E$ of (\ref{eq:Ex}) is proportional to $1/\Xi^6$ and so has maxima where $\Xi$ has minima.  We have already seen that in the $r\to\infty$ limit, $\Xi\to \xi$ of (\ref{xiInftyDefn}) whose minima lie on the spiral (\ref{curveLargeR}), which is the large-$r$ approximation to (\ref{curve}).  It can also be shown that the minima of $\Xi$ lie on the spiral (\ref{curve}) at all $r$.

Recall that our derivation of (\ref{curve}) in Section II was purely geometrical, relying only on the fact that the radiation is emitted tangentially, in the direction of the velocity vector of the quark and the fact that the pulse of radiated energy density propagates at the speed of light without spreading.  We have seen in Fig.~\ref{ekgplot} that at strong coupling and in the limit of a large number of colors, the radiation emitted by a rotating quark does indeed propagate outwards at the speed of light, without spreading.  This justifies the application of the geometrical arguments of Section II to the strongly coupled radiation.   The agreement with (\ref{Deltascaling}) and (\ref{curve}) then implies that the strongly coupled radiation is also emitted in the direction of the velocity vector of the quark: if it were emitted in any other direction, $\Delta$ would have the same $\gamma$-scaling as $\alpha$ and $\delta\theta$; if the radiation were, for example, emitted perpendicular to the direction of motion of the quark, the spiral of energy density would have the shape (\ref{WrongSpiral}) instead of (\ref{curve}) --- see Fig.~\ref{spiraldiagram}.  We therefore reach the following conclusions: at both weak and strong coupling, the energy radiated by the rotating quark is beamed in a cone in the direction of the velocity of the quark with a characteristic opening angle $\alpha\sim 1/\gamma$; at both weak and strong coupling, the synchrotron radiation propagates outward in a spiral with the shape (\ref{curve}), with pulses whose width $\Delta\sim 1/\gamma^3$ does not broaden.\footnote{Note that if the radiation is isotropic in the instantaneous rest frame of the quark, then in the inertial ``lab'' frame it will be beamed in a cone with opening angle $\sim 1/\gamma$ pointed along the velocity of the quark.     This, together with the result that the pulses of radiation propagate without broadening, would yield a spiral with the shape (\ref{curve}).
}

\begin{figure}[t]
\includegraphics[scale=0.3]{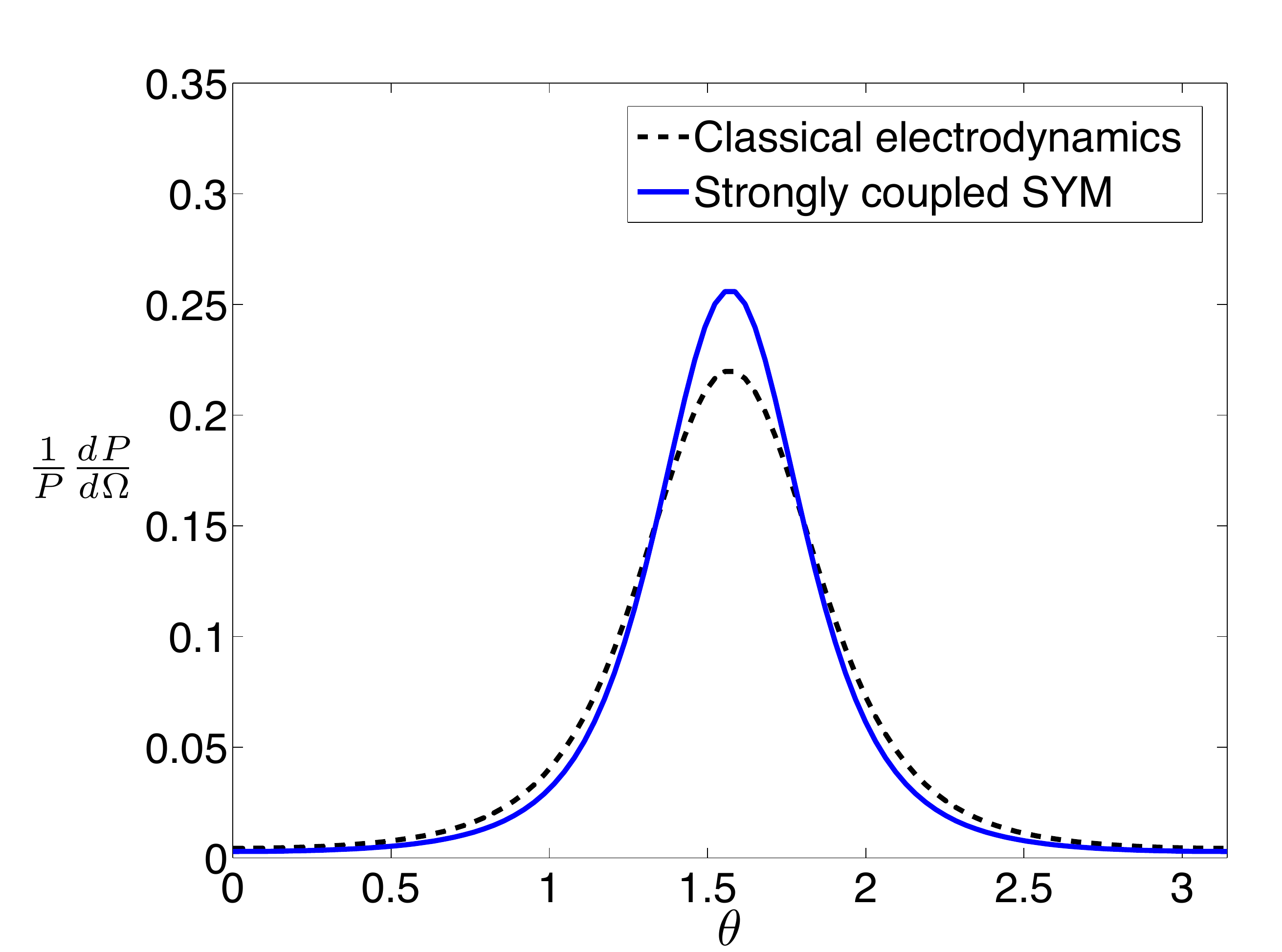}
\caption
    {\label{powerplot}
   The normalized time-averaged angular distribution of power radiated to large distances by a quark moving in a circle with velocity $v=0.9$  in classical electrodynamics and in strongly coupled $\mathcal N = 4$ SYM theory.   This normalized distribution is identical in weakly and strongly coupled ${\mathcal N}=4$ SYM theory.
In classical electrodynamics and in both weakly and strongly coupled ${\cal N}=4$ SYM theory, the time averaged angular distribution
   of power is localized about $\theta = \pi/2$ with a characteristic width $\delta \theta \sim 1/\gamma$.  
    }
\end{figure}

We now turn to the time-averaged angular distribution of power radiated 
through the sphere at infinity.  By time-averaging, we eliminate all dependence on azimuthal angle  but nontrivial dependence on the polar angle $\theta$ remains.  
Fig.~\ref{powerplot} shows the normalized time-averaged angular distribution of power in classical 
electrodynamics (\ref{vector}) and strongly coupled SYM (\ref{dpdOmega}) for velocity $v = 0.9$.  As is evident from the figure, in both
classical electrodynamics and $\mathcal N = 4$ SYM the power is localized about $\theta = \pi/2$ with a width $\delta \theta \sim 1/\gamma$.  The angular distribution of 
power is slightly more peaked for strongly coupled $\mathcal N = 4$ SYM.

Last, it is amazing to compare the time-averaged angular distribution of power in strongly coupled $\mathcal N = 4$ SYM, (\ref{dpdOmega}), to that in weakly coupled $\mathcal N = 4$ SYM, (\ref{VectorPlusScalar}).
The two results are identical, up to an overall normalization. Not only is there no broadening or isotropization, the pattern of radiation is identical at strong coupling to that at weak coupling!
Just as for the total power radiated, the angular distribution at strong coupling (\ref{dpdOmega}) 
is related to that at weak coupling (\ref{VectorPlusScalar}) by making the substitution (\ref{CorrectSubstitution}) in their prefactor.
We know that  the small difference between the normalized time-averaged angular distribution of power in classical electrodynamics and in weakly coupled $\mathcal N = 4$ SYM theory seen in Fig.~\ref{powerplot} comes from the fact that a quark in $\mathcal N = 4$ SYM radiates scalars as well as vectors.  It seems evident that the same is true for radiation in strongly coupled $\mathcal N =4$ SYM  theory as well.
Furthermore, in any strongly coupled conformal quantum field theory with a classical dual gravity description, the time-averaged angular distribution of power is given by our strongly coupled ${\cal N}=4$ SYM theory result 
(\ref{dpdOmega}).  
  It is remarkable how quantitatively similar radiation is at weak and strong coupling.


\subsection{Relation with previous work}

Next,  we return to the discussion of the relation between our present work and previous
works that we began in Section~I.  
In Ref.~\cite{Hofman:2008ar}, states produced by the decay of off-shell
bosons in strongly coupled conformal field theories were studied.  
It was shown in Ref.~\cite{Hofman:2008ar} that the event-averaged angular distribution 
of power radiated from the decay is isotropic (in the rest frame of the boson where the total momentum vanishes) 
and independent of the boson's spin.  This should be contrasted with similar weak
coupling calculations in QCD, where the boson's spin imprints an 
``antenna" pattern on the distribution of radiation at infinity \cite{Basham:1978zq, Basham:1978bw, Basham:1977iq}.  
Our results are not inconsistent with those of Ref.~\cite{Hofman:2008ar}.  First of all, we and the authors of Ref.~\cite{Hofman:2008ar} are considering very different initial states.  Second, as 
discussed in Ref.~\cite{Hofman:2008ar},  if their initial off-shell boson has a non-trivial distribution of momentum, 
this distribution imprints itself on the angular distribution of radiation at infinity and generically results in an anisotropic distribution of power.  
It is only the spin of the boson
which doesn't produce any anisotropy.

In Ref.~\cite{Hatta:2008tx} it was argued
that isotropization must occur in the strongly coupled limit due to parton branching.
Heuristically, the authors of Ref.~\cite{Hatta:2008tx} argued that at strong coupling
parton branching is not suppressed and that successive branchings can 
scramble any initially preferred direction in the radiation as it propagates out to spatial infinity.
However, our results clearly demonstrate that this need not be the case.  We see
no evidence for isotropization of the radiation emitted by the rotating quark.  In fact, the angular dependence of the time-averaged angular distribution
of power is the same at weak and strong coupling.  And, just as for weakly coupled radiation, the strongly coupled radiation that we have illustrated in Figs.~\ref{blingbling} and \ref{ekgplot} propagates outward in pulses that do not broaden: their extent in azimuthal angle at a given radius remains the same out to arbitrarily large radii and, correspondingly, their radial width is also unchanging as they propagate.   It is not clear in any precise sense what parton branching would mean in our calculation if it occurred, since no partons are apparent. But, the essence of the effect is the transfer of power from shorter wavelength modes to longer wavelength modes.  In our calculation, that would correspond to the pulses in Fig.~\ref{ekgplot} broadening as they propagate.  This we do not see.

\subsection{A look ahead to nonzero temperature}

It is exciting to contemplate shining a tightly collimated beam of strongly coupled synchrotron radiation into a strongly coupled plasma.   Although this will have its failings as a model of jet quenching in the strongly coupled quark-gluon plasma of QCD, so too do all extant strongly coupled approaches to this complex dynamical problem.  The calculation that is required is simple to state: we must repeat the analysis of the radiation emitted by a rotating quark, but this time in the AdS black hole metric that describes the strongly coupled plasma that arises at any nonzero temperature in any conformal quantum field theory with a dual gravitational description.    The calculation will be more involved, and likely more numerical, both because the temperature introduces a scale into the metric and because to date we have only a numerical description of the profile of the rotating string~\cite{Fadafan:2008bq}.  If we choose $T\ll 1/R_0$ and $\gamma\gg 1$ then the criterion (\ref{StirringCriterion}) will be satisfied and the total power radiated will be as if the quark were rotating in vacuum~\cite{Fadafan:2008bq}.  Based upon our results, we expect to see several turns of the synchrotron radiation spiral before the radiation has spread to radii of order $1/T$ where it must begin to thermalize, converting into hydrodynamic excitations moving at the speed of sound, broaden, and dissipate due to the presence of the plasma.  We will be able to investigate how the length scale or scales associated with these processes vary with $R_0$ and $\gamma$.  
Note that the typical transverse wavelengths of the quanta of radiation in the pulse will be of order $R_0/\gamma$ while their typical longitudinal wavelengths and inverse frequencies will be of order $R_0/\gamma^3$, meaning that we will have independent control of the transverse momentum and the frequency of the gluons in the beam of radiation whose quenching we will be observing.

Since at nonzero temperature the coils of the rotating string only extend outwards 
to some $\gamma$-dependent $r_{\rm max}$, it is natural to expect that this $r_{\rm max}$ will correlate with at least one of the length-scales describing how the spiral of synchrotron radiation shines through the strongly coupled plasma.   Our guess is that $r_{\rm max}$ will prove to be related to the length scale at which the energy carried by the nonhydrodynamic spiral of synchrotron radiation as in vacuum converts into hydrodynamic waves.   If so, we can estimate the parametric dependence of $r_{\rm max}$ at large $\gamma$, as follows. 
For $r\gg r_{\rm max}$, the power 
$P\propto \sqrt{\lambda} a^2 \propto \sqrt{\lambda} \gamma^4 /R_0^2$ 
radiated by the rotating quark will be carried by long wavelength hydrodynamic modes whose energy density will fall off $\propto \sqrt{\lambda}\gamma^4/r^2$.  For $r\ll 1/T$, we will have a spiral of synchrotron radiation whose energy density is $\varepsilon/r^2$ with $\varepsilon$ given by (\ref{farzoneeps}). Recalling from Section II that on the spiral $\xi=1-v\propto 1/\gamma^2$, we see from (\ref{farzoneeps}) that, on the spiral, the energy density is  $\propto \sqrt{\lambda}\gamma^8/r^2$.  We expect that for $1/T \ll  r \ll r_{\rm max}$ this energy density will be attenuated by a factor $\propto\exp(-{\rm const}\,r \,T)$.  The parametric dependence of the $r_{\rm max}$ at which the hydrodynamic modes take over from the nonhydrodynamic spiral can then be determined by comparing the parametric dependence of the two energy densities:
\begin{equation}
\frac{\sqrt{\lambda}\gamma^4}{r_{\rm max}^2} \sim \frac{\sqrt{\lambda}\gamma^8}{r_{\rm max}^2}  \exp\left(-{\rm const}\,  r_{\rm max} \,T\right)  \ , 
\end{equation}
meaning that 
\begin{equation}
r_{\rm max}\propto \frac{\log \gamma}{T}\ .
\end{equation}
Interestingly, the numerical results of Ref.~\cite{Fadafan:2008bq} do indicate that  $r_{\rm max}$ diverges slowly as $\gamma\to \infty$.

\section{Acknowledgments}

We acknowledge helpful conversations with Diego Hofman, Andreas Karch, Dam Son, Urs Wiedemann, Laurence Yaffe and Dingfang Zeng.  We thank Rolf Baier for pointing out a factor-of-two error in (\ref{scalar}), the power radiated in scalars in weakly coupled ${\cal N}=4$ SYM theory, in a previous version of this paper; 
only after correcting this error did it become evident that the angular distribution of power plotted in Figure 8 is identical in weakly and strongly coupled ${\cal N}=4$ SYM theory.  
This error was also corrected in Refs.~\cite{Hatta:2011gh,Baier:2011dh} which appeared after this paper was published. 
The work of PC is supported by a Pappalardo Fellowship in Physics at MIT. 
The work of HL is supported in part by a U.~S.~Department of Energy (DOE) Outstanding Junior Investigator award.
This research was
supported in part by the DOE Offices of Nuclear and High Energy
Physics under grants \#DE-FG02-94ER40818,  \#DE-FG02-05ER41360 and 
\#DE-FG02-00ER41132.
When this work was begun, PC was at the University of Washington, Seattle 
and his work was supported in part 
by DOE grant \#DE-FG02-96ER40956 and DN was at MIT and his work was
supported in part by the German Research
Foundation (DFG) under grant number Ni 1191/1-1.

\bibliographystyle{utphys}
\bibliography{refs}

\end{document}